\documentclass[onecolumn]{aastex63}

\usepackage{rotating}
\usepackage[utf8]{inputenc}
\usepackage[a4paper, total={6in, 8in}]{geometry}
\usepackage{natbib}
\usepackage{longtable}
\usepackage[page]{appendix}
\newcommand{\ksec}[1]{\, \\ \textit{#1 \\ \, \\}}

\usepackage{amsmath}
\usepackage{bm}
\usepackage{mathtools}

\usepackage{graphicx}
\usepackage{esvect}

\def\aap{Astro. Astrophys.}
\def\apjl{Astrophys. J. Let.}

\def\apjs{Astrophys. J., Supp.}

\def\deg{$^\circ$}

\shorttitle{Kepler Multi-Transiting Systems Properties}
\shortauthors{Judkovsky et al.}

\begin{document}
\received{May 22, 2023}

\title{Kepler Multi-Transiting Systems Physical Properties and Impact Parameter Variations}

\correspondingauthor{Yair Judkovsky}
\author[0000-0003-2295-8183]{Yair Judkovsky}
\affiliation{Weizmann Institute of Science, Rehovot 76100 Israel}
\email{yair.judkovsky@weizmann.ac.il}

\author[0000-0002-9152-5042]{Aviv Ofir}
\affiliation{Weizmann Institute of Science, Rehovot 76100 Israel}

\author[0000-0001-9930-2495]{Oded Aharonson}
\affiliation{Weizmann Institute of Science, Rehovot 76100 Israel}
\affiliation{Planetary Science Institute, Tucson, AZ, 85719-2395 USA}


\graphicspath{{./}{Figures/}}


\def\Nnewmasses{46 } 
\def\NnewmassesWithThreeSigma{41 } 
\def\Nmasses{101 }
\def\NmassesWithThreeSigma{95 }

\begin{abstract}
We fit a dynamical model to Kepler systems that contain four or more transiting planets using the analytic method \texttt{AnalyticLC}, and obtain physical and orbital parameters for \Nmasses planets in 23 systems, of which \NmassesWithThreeSigma are of mass significance better than $3\,\sigma$, and \Nnewmasses are without previously reported mass constraints nor upper limits. In addition, we compile a list of 71 KOIs that display significant Impact Parameter Variations (TbVs), complementing our previously published work on two- and three-transiting planet systems. Together, these works include the detection of significant TbV signals of 130 planets, which is, to our knowledge, the largest catalog of this type to date. The results indicate that the typical detectable TbV rate in the Kepler population is of order $10^{-2}\,\rm yr^{-1}$, and that rapid TbV rates ($\gtrsim0.05\,\rm yr^{-1}$) are observed only in systems that contain a transiting planet of an orbital period less than $\sim20$ days. The observed TbV rates are only weakly correlated with orbital period within Kepler's $\lesssim 100$ days-period planets. If this extends to longer periods, it implies a limit on the utility of the transit technique for long-period planets. The TbVs we find may not be detectable in direct impact parameter measurements but rather are inferred from the full dynamics of the system, encoded in all types of transit variations. Finally, we find evidence that the mutual inclinations distribution is qualitatively consistent with the previously suggested AMD  (angular momentum deficit) model using an independent approach.
\end{abstract}


\section{Introduction} \label{sec:Introduction}

The Kepler space telescope \citep{BoruckiEtAl2010}, utilizing the transit method, has brought a major increase in the number of known exoplanets. Following the path of  Kepler, the number of transit detections is rapidly increasing with the current Transiting Exoplanet Survey Satellite \citep[TESS,][]{RickerEtAl2010} and the future PLAnetary Transits and Oscillations of stars \citep[PLATO,][]{Rauer2014} missions.

The vast amount of photometric data calls for interpretation methods that enable us to extract the most out of the data accurately and efficiently. One of the most direct ways to extract planetary physical and dynamical information from photometric (and other types of) observational data is to run multiple $N$-body integrations. Such integrators were implemented at high accuracy \citep[e.g. ][]{WisdomHolman1991,Chambers1999,DeckAgolHolmanNesvorny2014}. The drawback of using $N$-body integrations is the relatively high computational cost, which increases with the time span of the data, a consideration that becomes more and more significant as the available data increases and is now frequently spanning multiple decades.
In order to enable a more efficient interpretation of photometric data, different authors approached the interpretation process in an analytic manner, usually using Transit Timing Variations \citep[TTVs, ][]{AgolSteffenSariClarkson2005}. When TTVs are detectable, the orbital configuration and the planetary masses may be estimated. This made TTVs a tool of great utility and thus a topic of thorough investigation, a few examples being: \citet{NesvornyMorbidelli2008,Nesvorny2009,NesvornyBeauge2010}, a series of papers that analyzed the TTV problem and developed the method \texttt{ttvim}; \citet{LithwickXieWu2012} who analyzed the fundamental TTV mode to first order in eccentricity; \citet{DeckAgol2015} who highlighted the synodic "chopping" TTV, \citet{HaddenLithwick2016} who extended the method of \citet{LithwickXieWu2012} to second order in eccentricity, and \citet{AgolDeck2016} who published the publicly available code \texttt{TTVFaster} that analytically calculates the TTV to first order in eccentricity. All of the above, however, analyze the TTVs - which are only a by-product of the main measured quantity: the stellar flux.
There are two main drawbacks to performing a TTV analysis rather than a full light curve fitting. First, for small planets with shallow transits, an individual transit (unlike the full "folded" light curve) may not be significant in the data, making timing severely limited or impossible, thus preventing TTV analysis. In such cases, a full, global, light curve model can extract information that cannot be extracted by attempting to fit transit times. Second, restricting the analysis to TTV ignores other types of transit variations, such as Transit Duration Variations (TDV). These are of high interest in probing forces out of the plane but are unfortunately even more difficult to measure than TTVs on long cadence data such as Kepler's; the utility of such information, in the rare cases where it is observable, can be seen in  the detection of the mutual inclination between planets b and c in the Kepler-108 system \citep{MillsFabrycky2017}. This mutual inclination was later explained to have potentially arisen from a resonant interaction with a binary member of the host star, called ivection resonance \citep{XuFabrycky2019}; thus, in this case, the detection of TDV eventually led to the understanding of a formerly unrecognized dynamical phenomenon.

In a previous study \citep[][hereon paper I]{Judkovsky2022a}, we introduced \texttt{AnalyticLC}, a method and code implementation for modeling light curves. This analytic method balances accuracy and efficiency and can also model RV and astrometry to enable joint fitting of these data types. We used this method to analyze a set of two- and three-transiting planet systems \citep[][hereon paper II]{Judkovsky2022b}. This work led to 140 new planetary masses and orbital properties, along with a list of KOIs that display Transit Impact Parameter Variations (TbV). Statistics of such transit variations are important for characterizing the distribution of planetary systems properties, offering opportunities to derive inferences on population properties \citet{FabryckyEtAl2014}, \citet{XieEtAl2016}, \citet{Millholland2021}.

In this work, we extend the study to systems that contain four or more transiting planets. Such systems are interesting due to the dynamical richness embedded in many planetary interactions involving multiple resonances. Multi-planet interactions can also lead to phenomena that are more complex than direct pairwise interactions. An example of such a scenario was presented in the KOI-500 system by \citet{Ford2011}, who mentioned that the similarity between the distances from resonance might result in a dynamical interplay that could be manifested in the observations. In paper I, we formulated this kind of interaction and denoted it Super-Mean-Motion-Resonance (SMMR). This dynamical phenomenon arises from the simultaneous interaction of three planets in a near-resonant chain. In multi-transiting planet systems, such effects are more likely to be significant. Here we report three such cases in KOI-707, KOI-1589, and KOI-2038 (see \S\ref{sec:Individual}).

This paper is organized as follows. In \S\ref{sec:Methods} we review the methods used in this work. In \S\ref{sec:Results} we review the main findings from this work in two aspects: planetary masses and statistics of measured TbVs. In \S\ref{sec:Individual} we refer to each of the analyzed systems, including comparison to former literature results and discussing the main dynamical features of each system. In \S\ref{sec:Discussion} we conclude and discuss future prospects. Numerical results are provided in tables in \ref{appendix}.

\section{Methods}\label{sec:Methods}

The methods used in this work are the same ones used in paper II, with a slight adjustment: The number of walkers in the DE-MCzs process \citep[Differential Evolution Monte Carlo with a sampling of past states using snooker updates,][]{BraakVrugt2008} is larger because the number of planets, and the corresponding number of model parameters, is larger. We repeat the main stages below.

\subsection{Data Reduction}

The detrended light curve is obtained by applying a cosine filter to the Presearch Data Conditioning (PDC) Maximum A Posteriori (MAP) Kepler data \citep{Stumpe2014}. The cosine filter is based on the one used by \citet{Mazeh2010}, and modified such that the shortest time scale is four times the transit duration time, as provided by NASA Exoplanets Science Institute\footnote{\url{https://exoplanetarchive.ipac.caltech.edu/}} (NExScI). The aim of this choice is to avoid the filter affecting the transit shape itself. Outliers beyond $5\,\sigma$ were iteratively rejected (outliers were removed separately for Kepler long and short cadence data).

The raw data detrending and the model fitting affect each other, so we performed an iterative process: having detrended the light curve, we fitted a model, normalized the raw data by the best-fitting model, and then detrended again. This iterative process was regarded as converged when the $\chi^2$ score did not improve from the previous iteration by more than the equivalent of $1\,\sigma$, which is defined as the 68th percentile of the $\chi^2(N)$ distribution, where $N$ is the number of model degrees of freedom. This process has already been proven robust in the former study paper II, and its robustness was reaffirmed here.

We used all available Kepler data (long and short cadence). Binning was applied for the long cadence data points, where the number of binning points was set by requiring that the error caused by the binning is an order of magnitude smaller than the typical data uncertainty, using the formulae given by \citet{Kipping2010a}.

The model function to evaluate the instantaneous flux was \texttt{AnalyticLC}, which was described in paper I and was already used to study Kepler's two- and three-transiting planets systems in paper II. The model uses an analytic 3D dynamical model coupled with the flux calculation formulae of \citet{MandelAgol2002}. For flux modeling, we took the quadratic limb darkening parameters from NExScI.

\subsection{Model Parameters}\label{sec:ModelParameters}

The model includes eight parameters per transiting planet. Two relate to the physical properties of the planet: $\mu$ (planet to star mass ratio) and $R_{\rm p}/R_*$ (planet to star radius ratio). The other six are (equivalent to) the orbital elements: $P$ (orbital period), $T_{\rm mid0}$ (reference time of mid-transit), $\Delta e_x, \Delta e_y$ (eccentricity vector components differences from the neighboring interior planet; for the innermost planet these are just the eccentricity vector components), $I_x=I\cos{\Omega}, I_y=I\sin{\Omega}$ (inclination vector components, where $I$ is the inclination with respect to the $z$ axis and $\Omega$ is the longitude of ascending node). The axis system is defined such that the $x$ axis points from the center of the star to the observer, $y$ lies on the sky plane such that the innermost planet crosses the sky plane on it, and the $z$ axis lies on the sky plane and perpendicular to the $x$ and $y$ axes. Using this axis system, $e_x=e\cos{\varpi}$ is the eccentricity component on the line of sight and is positive if the periapse is between the star and the observer, where $e$ is the eccentricity magnitude, and $\varpi$ is the longitude of periapse. A diagram illustrating the orbital elements and the axis system is shown in Figure~1 of paper II.

For the innermost planet, we fix $\Omega=\pi/2$, {\it i.e.} we set $I_x=0$. An additional model parameter is the innermost planet semi-major axis to stellar radius ratio, $a/R_*$ \citep[as in ][]{OfirDreizler2013}, leaving it still with eight parameters. Using Kepler's law, the semi-major axis of the other planets in the system is then scaled from the innermost one.

\subsection{Non-Linear Fitting Method}\label{sec:NonLinearFitting}

We use our code implementation of the Differential-Evolution Markov-Chain with Snooker Update \citep[]{BraakVrugt2008} to generate a posterior distribution and obtain a best-fitting model. This method suits multi-dimensional problems with correlated parameters, as in our case of fitting a dynamical model to a light curve.

We limited the eccentricity components magnitude to smaller than 0.6 and the inclination components magnitude to larger than 50\deg{} since the model is usually invalid beyond these relatively extreme configurations; in practice, the solutions usually converged to much smaller values of a few percent/a few degrees.

The choice of the number of walkers was guided by the value of $N\log{N}$ where $N$ is the number of fitted parameters. There is no general prescription for the optimal number of walkers; we based this choice on experimenting with different types of non-linear problems and on its satisfactory performance is our previous work on two- and three-transiting planets  (paper II). Convergence was declared upon reaching a Gelman-Rubin criterion \citep{Gelman1992,Brooks1998} of less than 1.2 for all parameters over 10,000 generations. After the removal of the burn-in, we were left with a well-mixed sample, from which statistical inference of the parameters was made. The parameter values described in the tables in \ref{appendix} are the sample median and the difference between the median and the 15.865 and the 84.135 percentiles, corresponding to $1 \,\sigma$ in a normal distribution.

The walkers' initial states in orbital periods, reference times of mid-transit and planet-to-star radius ratio, and planet-to-star mass ratio are distributed based on values and errors from NExScI.  If no planetary mass is provided, we translate the archive planetary radius to an initial guess on mass by using an empirical formula given by \citet[][Fig.~8]{Weiss2018}.

We ran the non-linear fitting procedure five times per system in order to check if it converges to the same minimum consistently, using a different realization of the initial walkers' distribution at each run. In our former study of two- and three-transiting planet systems, we empirically found that this number suffices for most cases for obtaining repeatable results. We also note that each run contains 112 walkers (for four-planet systems), 144 walkers (for five-planet systems), and 184 for six-planet systems. Multiplied by tens of thousands of generations at each run, the generated samples covered a large fraction of the parameters space.
In \S~\ref{sec:Individual} we discuss the results system-by-system and note cases with ambiguous solutions that require further investigation.
 
The initial guess of $a^{(1)}/R_*$ was obtained from the literature stellar mass and planetary orbital period; whenever possible (for the decisive majority of planets), we used stellar data from \citet{Berger2020}. For KOI-593 and KOI-1422, where this source had no data, we used NExScI.

The fit parameters of \texttt{AnalyticLC} are dimensionless; we translate them to absolute masses and radii using stellar literature data from \citet{Berger2020}.

\subsection{Verification vs. N-body Integration}\label{sec:Nbody}

In order to verify the validity of the best-fitting analytic model generated by \texttt{AnalyticLC}, which is calculated using truncated series expansion of the full 3D gravitational interaction,  we compare it to a model generated by an N-body integration using \texttt{Mercury6} \citep{Chambers1999}. Following \citet{OfirEtAl2018}, the data uncertainties normalize the flux differences between the analytic model and the N-body model, and the sum of the squares of these normalized differences provides a $\chi^2$-like estimate of the mismatch between the analytic and the N-body model at the best fitting parameters set:

\begin{equation}
    \chi^2_{\rm Nbody}=\sum_{t}\frac{(F_{\rm ALC}(t)-F_{\rm Nbody}(t))^2}{dF(t)^2},
\end{equation}

where $F_{\rm ALC}(t)$ and $F_{\rm Nbody}(t)$ are the flux values obtained from \texttt{AnalyticLC} and the N-body integration at each data point, respectively, and $dF(t)$ is the data uncertainty of each data point. $\chi^2_{\rm Nbody}$ measures the ability to statistically discern between the two models given the data available.

We used the empirical CDF obtained from the posterior distribution to translate $\chi^2_{\rm Nbody}$ value to CDF by interpolation and then to the equivalent number of standard deviations assuming a normal distribution.
We refer to this as $\sigma_{\rm Nbody}$, which quantifies the systematic error of the model relative to the statistical error of the data.

\begin{figure}[h]
    \includegraphics[width=1\linewidth]{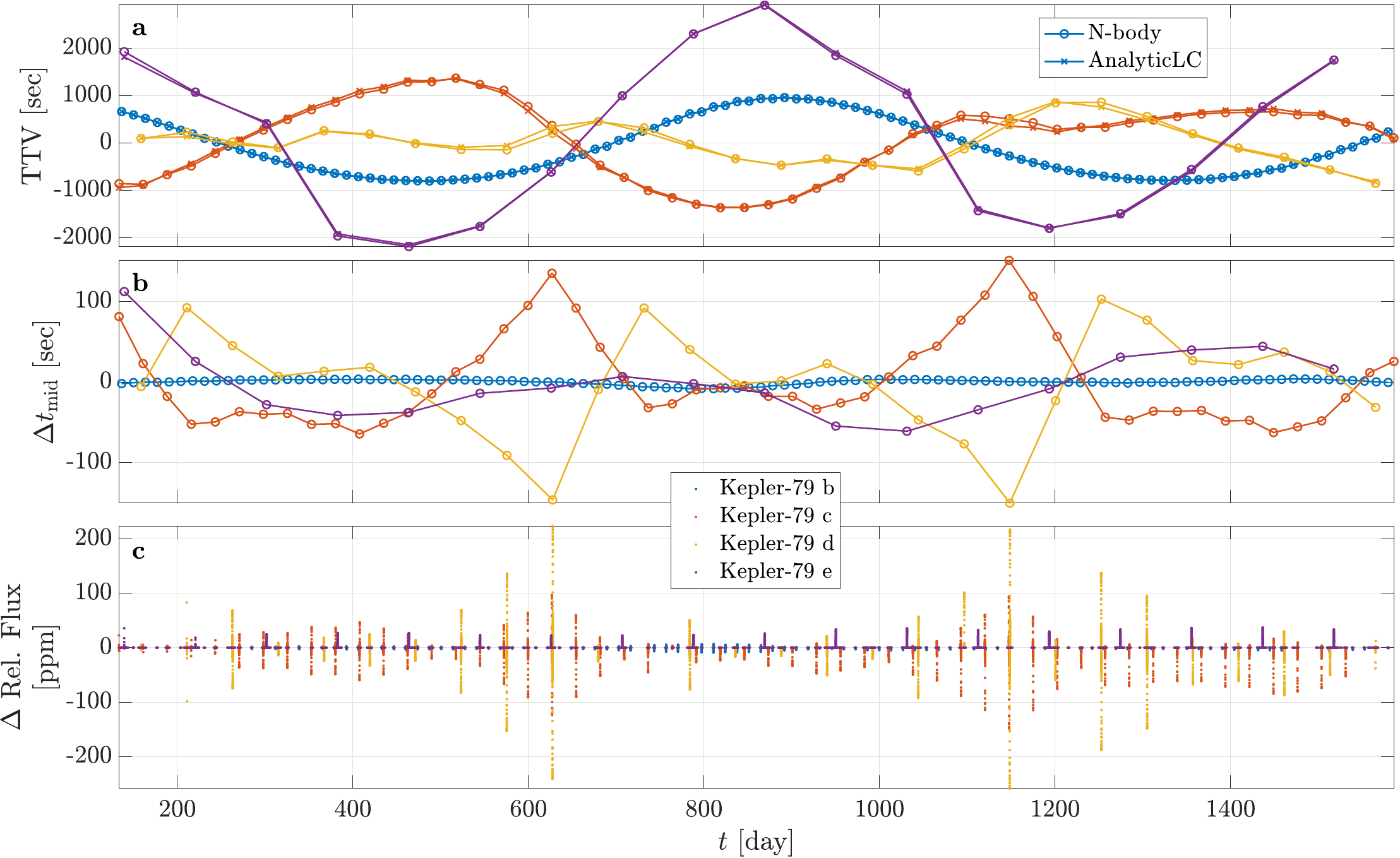}
    \caption{An example for the N-body matching test performed on the best fitting solution of the planetary system of Kepler-79 (KOI-152), demonstrating the ability of \texttt{AnalyticLC} to generate a model consistent with full N-body integration of a four planets system. The mismatch is quantified as $\chi^2_{\rm Nbody}\sim10$,
  ($\sigma_{\rm Nbody}=0.0321$) - a systematic error much smaller than the statistical error arising from the data uncertainty. (a) TTV pattern of the N-body model (o) and the \texttt{AnalyticLC} model (x) for the four planets in this system (b in blue, c in red, d in yellow, e in purple). The symbols are on top of each other at this scale. (b) "Residuals": Times-of-mid-transit mismatch between the N-body generated model and \texttt{AnalyticLC}, which is of order a few seconds for the innermost planet and of order of a minute for the three outer planets. 
  (c) Manifestation of the mismatch to terms of relative flux. The mismatch standard deviation within the in-transit points is 0.7, 10, 25.5, and 9 ppm for the four planets correspondingly, innermost to outermost.  The typical {Kepler} short-cadence data (which is almost all of the data) uncertainty for this star ($\sim 1000-1200$ ppm).}
    \label{fig:NbodyTest}
\end{figure}

Figure~\ref{fig:NbodyTest} shows an example of the comparison between \texttt{AnalyticLC} and the results of a full N-body integration for the four-transiting planets {Kepler}-79 system (KOI-152). The best-fit solution of \texttt{AnalyticLC} for this system agrees well with an N-body integration ($\sigma_{\rm Nbody}=0.0321$), implying that the model error with respect to full N-body integration is two orders of magnitude smaller than the statistical error arising from the data uncertainty. By performing the computationally heavy multi-dimensional search using \texttt{AnalyticLC}, while validating only the best-fit point using an N-body integrator, we ensure both the efficiency of our search and the accuracy of our result.

\begin{figure}[h]
    \includegraphics[width=1\linewidth]{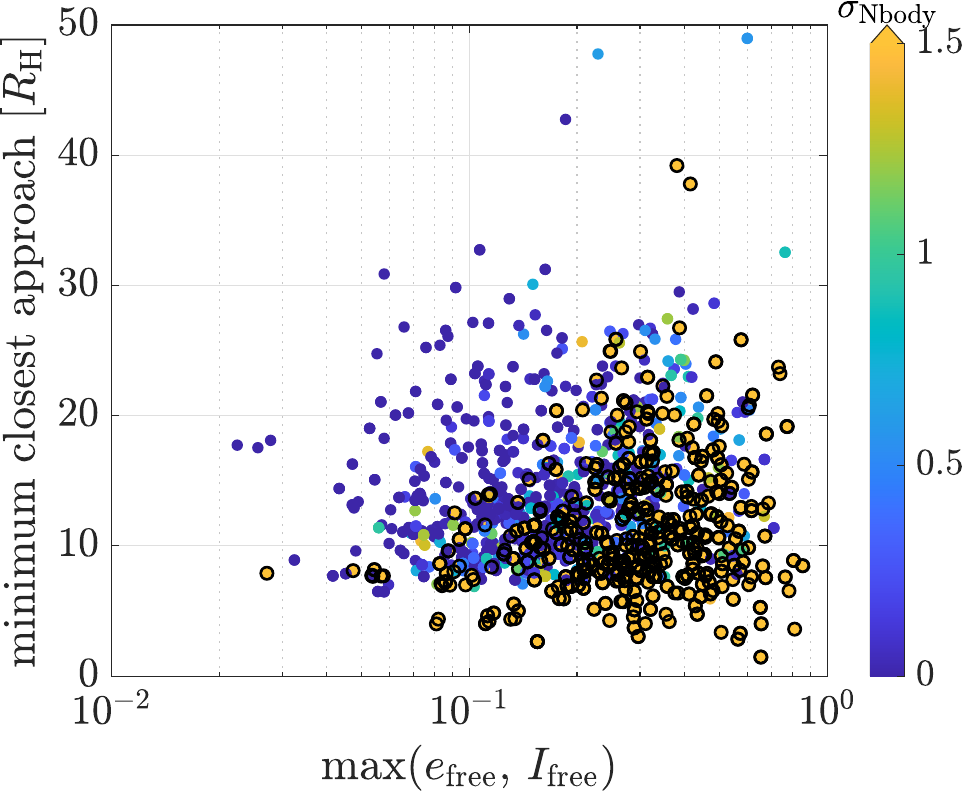}
    \caption{An overview of the N-body match with \texttt{AnalyticLC} for all our runs, including both this work and the former work of the two- and three-transiting planets systems, including both runs that did or did not pass our acceptance criteria. This map demonstrates the accuracy limits of \texttt{AnalyticLC}. The horizontal axis is the largest magnitude of all free eccentricities and inclinations in the system at the best-fitting solution derived by \texttt{AnalyticLC}. The vertical axis is the minimal closest approach in the system in units of mutual Hill radii.  The color of the points represents the value of $\sigma_{\rm Nbody}$, with points at which $\sigma_{\rm Nbody}>1.5$ (which do not pass our acceptance criterion) also marked  with black edges.}
    \label{fig:NbodyMatchAllSystems}
\end{figure}

In paper II we have shown a map of the validity of the models generated by \texttt{AnalyticLC}. Here we extend this map to include both the two- and three-transiting planets systems from that work and the systems from this current work. This map is shown in Figure~\ref{fig:NbodyMatchAllSystems}. Systems with eccentricities and inclinations of up to $0.1-0.2$ and orbital separations of $\gtrsim10$ mutual Hill radii can be adequately modeled by \texttt{AnalyticLC} in the vast majority of cases to the precision of the Kepler data. There are also systems of smaller separations and/or larger eccentricities in which the model is adequately correct. Systems that include close approaches closer than $\sim7-8 R_{\rm H}$ are usually not consistent with the N-body-integrator-based result. This is an expected outcome of our analytic approach, which inherently assumes slow variations of all orbital elements except the mean longitudes - an assumption that breaks when close encounters occur.

It is noteworthy that our analysis is a type of model fitting, and therefore it has some inherent limitations. For example, it will not model any real part of the system that was not included in the model, e.g. non-transiting planets.
It is clear from Fig. \ref{fig:NbodyMatchAllSystems} that \texttt{AnalyticLC} found that some systems have relatively high eccentricities - much higher than is typical for multi-planet systems. We believe that at least some of these systems actually contain additional, non-transiting planets that affect the dynamics of the system, and the observed eccentricities are simply a reflection of \texttt{AnalyticLC}'s attempt to use the degrees of freedom open to it to match the observed data.

Regarding the current sample of systems, from the 23 systems that passed all our tests (see below) and attain $\sigma_{\rm Nbody}<1.5$, for 16 systems $\sigma^2_{\rm Nbody}<0.1$, implying a very good match (such that when summing in quadrature the systematic and statistical errors, the systematic error contribution is an order of magnitude smaller than the statistical one). 

\clearpage

\subsection{Comparison to Strictly Periodic Circular Model}\label{sec:AlternativeModel}

Our model has eight parameters per planet, while a minimal model of a circular strictly periodic orbit would involve only 5 for the innermost planet ($P, T_{\rm mid0}, R_{\rm p}/R_*, a^{(1)}/R_*, b$, where $b$ is the impact parameter) and only 4 for the other planets ($a/R_*$ for the other planets can be deduced from the period ratios). In order to justify the larger number of parameters in the full model, we calculated the Akaike Information Criterion (AIC) \citep{Akaike1974} for both the dynamical and strictly periodic models, and we validate only solutions where the AIC of the dynamical model is better. A more detailed discussion of the selection of AIC as our information criterion is given in paper II.

\subsection{Solution Consistency with Stellar Parameters}

One of our model parameters is $a^{(1)}/R_*$, the ratio between the semi-major axis of the innermost planet and the stellar radius. This parameter can be obtained alternatively from literature values for the stellar mass and radius \citep{Berger2020,Fulton2018}, and the measured orbital period of the innermost planet. We checked the consistency of our converged solution with the literature-obtained value with respect to the literature error and our posterior distribution error on $a^{(1)}/R_*$. In the systems for which we provide a dynamical solution, there are no cases in which this difference is larger than $2.7\,\sigma$, where $\sigma$ is the root-sum-squares of the error estimates on $a^{(1)}/R_*$ from our posterior distribution and the literature. In fact, the differences between our estimated $a^{(1)}/R_*$ and literature values distribute nicely around zero with a standard deviation close to unity, resembling a normal distribution. We, therefore, consider our results in agreement with the literature's stellar parameters.

\subsection{Consistency among Solutions}\label{sec:Consistencty}

As described above, we ran DE-MCzs five times for each planetary system, each time with a different random seed. We discarded solutions for which our model is not compatible with a full N-body integration (see \S\ref{sec:Nbody}), and solutions for which the number of model parameters is statistically unjustified (see \S\ref{sec:AlternativeModel}). In addition, we discarded solutions that imply unreasonably high planetary density - more than $2\,\sigma$ above $12\,{\rm g}\,{\rm cm}^{-3}$, the approximate density at the base of Earth's outer core \citep[][chapter 2]{SOROKHTIN201113}. 
This process left us with a subset of runs for each KOI. If all of them converged to the same maximum-likelihood region in parameters space, the solutions are consistent with each other, and we report the obtained solution posterior distribution statistics. If they converged to different regions in parameters space, we report all of them and regard one of them as the "adopted" solution, recognizing this solution is not unique. Selection of the adopted solution is based on differences in fit quality ($\chi^2$). If there are a few solutions of similar fit quality, we publish all of them and select one of them as the adopted solution based on physical reasoning, e.g. favoring solutions with plausible planetary densities, with small eccentricities for short-orbit planets, and with smaller mutual inclinations, as we treat here systems with four or more transiting planets which are statistically likely to possess small inclinations \citep{RagozzineHolman2010}). For systems with more than one solution, we refer to the various solutions in \S\ref{sec:Individual}.

For eleven out of the 23 systems with valid dynamical solutions, we report one solution, either because one was than the others by more than $\sim3\,\sigma$ (e.g., KOI-152), or because only one solution passed all our validity criteria (e.g., KOI-720). In some cases, a few runs converged to the same local minimum, yielding consistent solutions; in such cases, we report it once. For twelve systems, we report more than one solution (five with two solutions, seven with more). 

In order to test the solutions' consistency, for each pair of runs we check the overlapping area between the Probability Density Function (PDF) of the posterior marginal distributions of each parameter. This overlapping area, denoted as $\eta$, is a measure of the similarity of two distributions \citep{Pastore2019}, defined for two PDFs of the parameter $x$, $f_1(x)$ and $f_2(x)$, as
\begin{equation}
    \eta = \int \min{[f_1(x),f_2(x)]}dx,
\end{equation}
where the integration is restricted to the range of validity of the variable $x$. A value of $\eta=1$ means that $f_1,f_2$ are the same distributions. To calibrate the values of this quantity, two Gaussians with the same standard deviation shifted from one another by one standard deviation yield $\eta\simeq 0.62$; two Gaussians shifted by two standard deviations yield $\eta\simeq 0.31$. Any value much smaller than that would mean the distributions are significantly different; in cases where at least one of the parameters $\eta \gtrsim 0.31$ we publish both solutions involved (as long as they have a similar fit quality and pass all the criteria detailed above). This is the case, for example, for KOI-1336.

\subsection{Sample of Fitted Planetary Systems}
Our former work (paper II) analyzed the data of two- and three-transiting planets systems. Here we selected systems of four or more transiting planets of Kepler.

Our initial sample consisted of 56 four-planet systems, 17 five-planet systems, and one six-planet system (KOI-157). This sums up to 74 systems for which we began the computation process. For 67 of these systems, at least one run converged to a final solution. The validity criteria detailed above were applied to these solutions, namely: (i) N-body matching (illustrated in Figure~\ref{fig:NbodyTest}); (ii) AIC improvement over a strictly-periodic model (\S\ref{sec:AlternativeModel}); and (iii) plausible planetary density (\S\ref{sec:Consistencty}). This process filtered out some of these systems. In the end, we were left with a sample of 23 systems (fifteen of four planets, seven of five planets, and one of six planets), consisting of \Nmasses planets with a dynamical solution, for which we reported masses and orbital elements.

We combine the error obtained from our posterior distribution of the planet-to-star mass ratio and the literature error on stellar mass in quadrature to obtain the total error on the absolute planetary mass. Similarly, we combine the error obtained from our posterior distribution of the planet-to-star radius ratio and the literature error on stellar radius in quadrature to obtain the total error in absolute planetary radius. In most cases, the majority of the error in mass stems from the fit and not the stellar parameters. For the radius, it is the other way around: the radius ratio is well-constrained by the fit, and most of the uncertainty arises from the uncertainty in stellar absolute radius.

\clearpage

\section{Results}\label{sec:Results}

\subsection{General}

\begin{figure}[h]
    \includegraphics[width=0.9\linewidth]{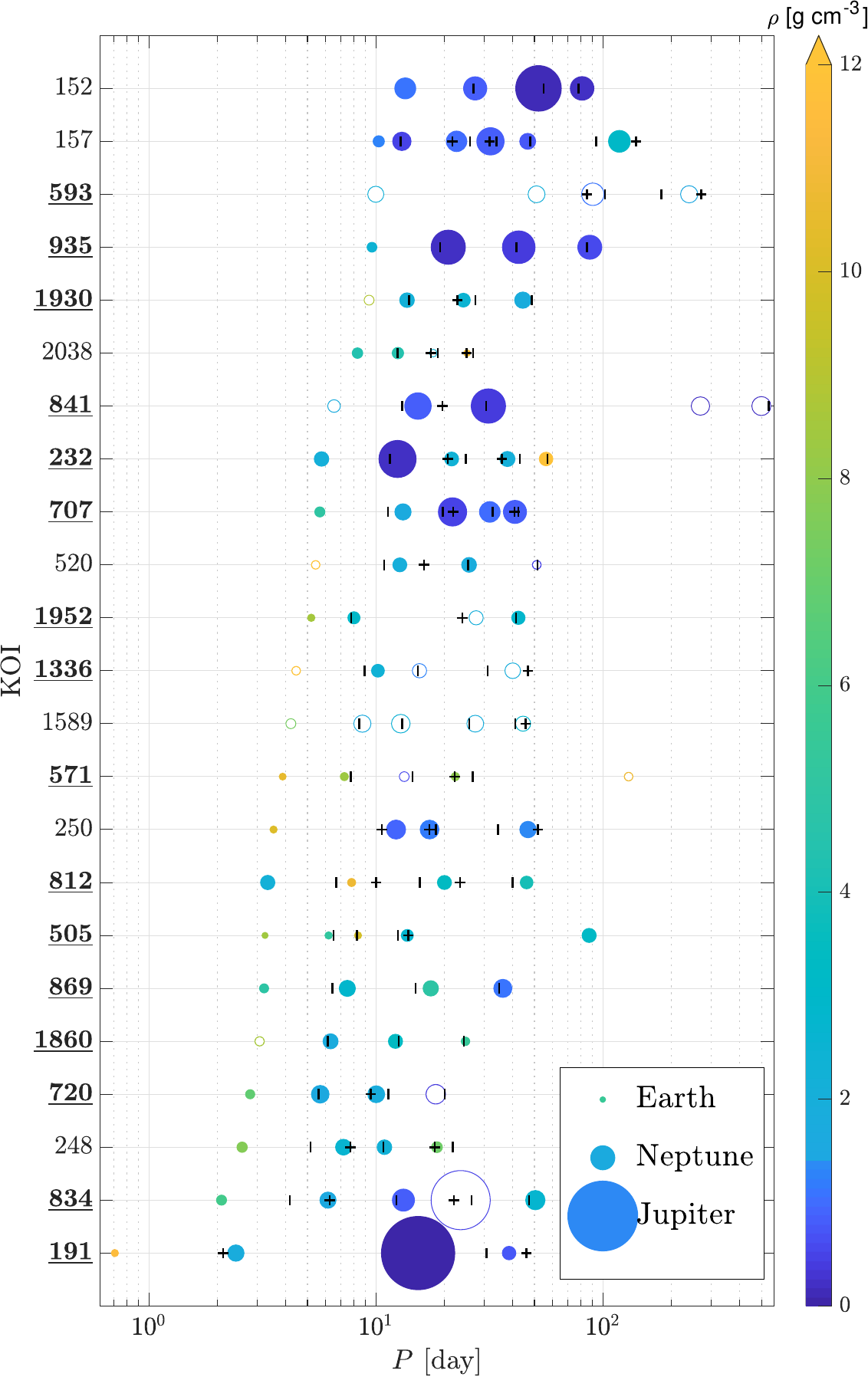}
    \caption{Orbital period, radii, and densities of the planets with mass estimates from this work. KOI numbers for which at least one planet does not have a previously reported mass value or upper limit are shown in underlined bold text. The size of the circles represents the absolute planetary size, and the color scale indicates our adopted density. Only planets with densities estimated to the significance of more than $4\,\sigma$ are color-filled; others are empty. For some of the planets (e.g. in the systems KOI-593 and KOI-1589) the relative masses are well-constrained, but the absolute mass (and hence density) is not well-constrained due to a high uncertainty on the stellar mass and radius, and hence these planets are shown here unfilled. For reference, the legend shows the size and density of Earth, Jupiter, and Neptune. The short vertical black lines indicate the locations of the closest first-order MMRs to the observed period ratio of adjacent planets. The black crosses similarly indicate the locations of second-order MMRs - note that many planets are close to these MMRs. Both first and second MMRs are indicated only if they are close to the observed period ratios. We do not show second-order MMRs which are a multiplication of a first-order MMR (e.g. 4:2 and 2:1).}
    \label{fig:FittedPlanetsOverview2}
\end{figure}

Of the 23 systems for which we found a valid solution, fifteen contain four transiting planets, seven contain five transiting planets, and one contains six transiting planets (KOI-157, Kepler-11), summing to \Nmasses planets in total. In Figure~\ref{fig:FittedPlanetsOverview2} we show a map of orbital periods, planetary radii and planetary densities obtained in this work, and resonances locations. This map gives a brief overview of the planetary systems for which physical properties were obtained in this work. Many systems are near at least one resonance location, as the proximity to resonances generates the high-amplitude TTVs that enable mass estimation. The uniform spacing and the similar planetary radii are visually apparent in many systems \citep[the so-called "Peas in a pod", ][]{Weiss2018a}. In some of these systems, this similarity also seems to occur in planetary masses and densities; this similarity has been recently studied quantitatively and has been shown to exist, though to somewhat a smaller extent than the radii similarity \citep{Otegi2022}. 

\clearpage

\subsection{New Constraints on Planetary Masses}

\begin{figure}[h]
    \includegraphics[width=1\linewidth]{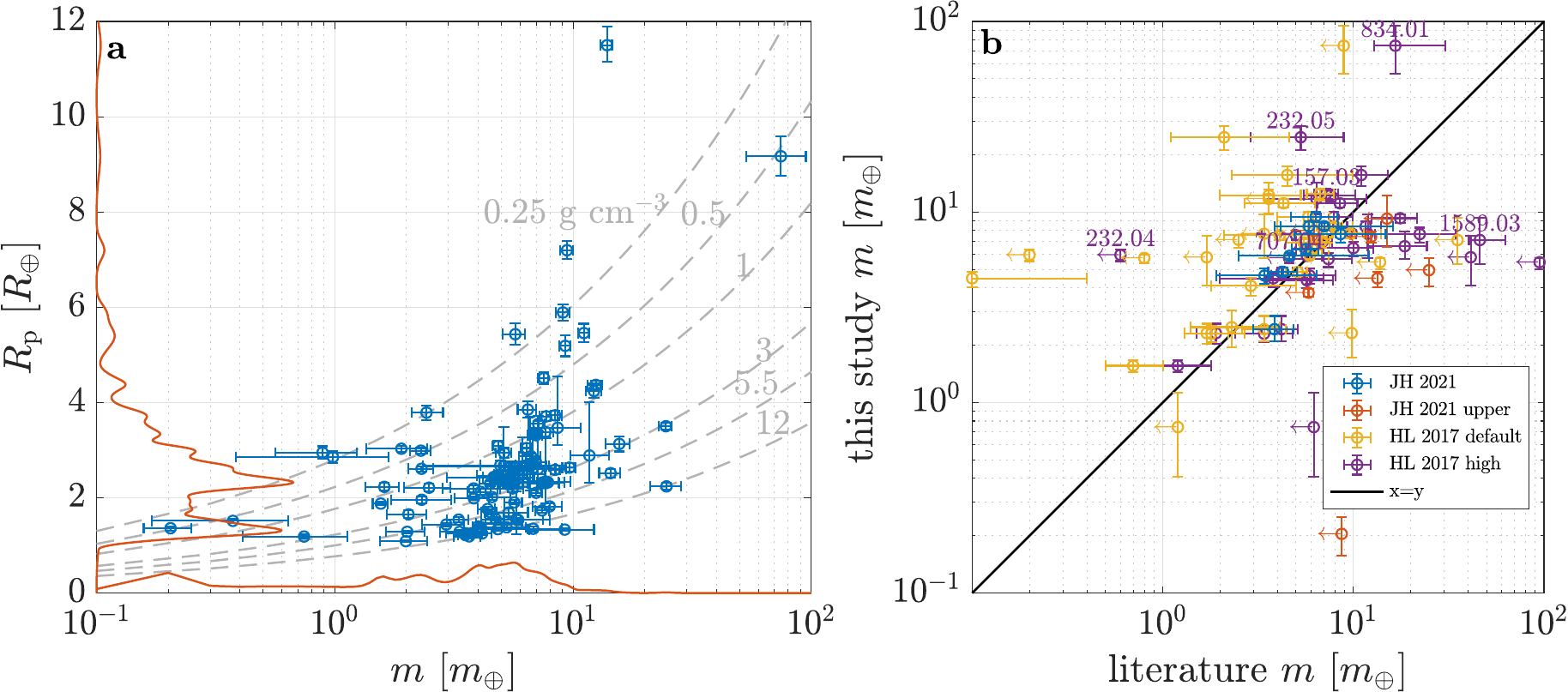}
    \caption{Planetary masses and radii obtained from this work. This figure shows the overall spread of masses and radii and the good agreement of planet masses with former literature values. (a) Mass-radius diagram. Each blue error bar is related to a single planet. The red lines are normalized histograms of the masses and radii obtained by summing up the PDFs of all points. As the sample is small and radii are well determined, the radii weighted histogram appeared more as a collection of discrete values, so it was smoothed using a Gaussian kernel of width $0.3\, R_\oplus$ to better show population-wide trends. The gray contours are constant-density curves. (b) Comparison of masses obtained in this work with literature values: JH21 values (blue) and upper limits (red); HL17 default mass prior, three of which show upper limits only (yellow) and high mass prior (purple). We omit planets for which literature values are within less than $2\,\sigma$ of zero. For six objects labeled with their KOI numbers, the results obtained in this work disagree with the results of HL17 by more than $2\,\sigma$; these are discussed in more detail in the text. The black line shows the identity function.\label{fig:FittedPlanetsOverview1}}    
\end{figure}

In the left panel of Figure~\ref{fig:FittedPlanetsOverview1} we show the mass-radius spread of the \Nmasses planets in these 23 systems, with curves of constant density and the one-dimensional marginal distribution of this planet population. The radius gap at $R_{\rm p} \sim 1.8 R_{\oplus}$ \citep{Fulton2017} is visually apparent; the typical planetary mass is $5-10\,m_\oplus$. The sample is dominated by planets with densities of 1-3$\,{\rm g}\,{\rm cm}^{-3}$. While these densities are definitely not consistent with a mostly rocky composition, they are also consistent with the Solar System's ice giants only at the lower end of this range. This suggests that the typical composition of these exoplanets is unlike anything found in the Solar System, with the total mass more evenly divided between the rocky core and volatile envelope. Some planets have densities lower than 1$\,{\rm g}\,{\rm cm}^{-3}$ - these are planets massive enough to keep their large gas atmosphere, and they are not extremely hot (none of them has an orbital period shorter than 10d). The uppermost point in this panel belongs to KOI-191.01; the derived mass and radius imply density of $\sim0.05\,{{\rm g}\,{\rm cm}}^{-3}$; as we discuss in \S\ref{sec:Individual} the inference on this planet is questionable and deserves further study. The next low-density planet belongs to KOI-152.01, with a planetary mass of $\sim9.4\,m_\oplus$ and a radius of $\sim7.2\,R_\oplus$; this value is more reliable and is consistent with past literature  to $\sim1.5\,\sigma$ \citep{JontoffHutter2014, HaddenLithwick2017, JontofHutter2021}.  The curve of 5.5$\,{\rm g}\,{\rm cm}^{-3}$ was chosen as it is the approximate Earth's bulk density; the curve of 12$\,{\rm g}\,{\rm cm}^{-3}$ was chosen as it is the estimated approximate density at the base of Earth's outer core \citep[][chapter 2]{SOROKHTIN201113}, and as such it is used to delineate the upper limit of accepted solutions (\S\ref{sec:Consistencty}).

In the right panel of the same figure, we show a comparison of the masses obtained in this work with values of HL17 and JH21, which are studies that have a large number of common KOIs with our work. 
Our results are in good agreement with the masses obtained by JH21 (some of which were given as upper limits; these are indicated by arrows pointing left).

 HL17 applied a default mass prior and a high mass prior and got two results for each planet; the masses we obtained are typically larger than their low-prior mass and are close to their masses obtained from high-mass priors. This is a reasonable outcome, as our prior is uniform in mass, the same as their high-mass prior. 
 
 For KOI-834.03, HL17 report a mass of $239.1\,m_\oplus$ which is out of the bounds of this plot; the value is not plausible as this is a $1.95\,R_\oplus$ planet candidate.
 For six planets, the results we obtained are different than the results obtained by HL17 by more than $2\,\sigma$. All of these cases are indicated in fig. \ref{fig:FittedPlanetsOverview1} specifying their KOI numbers: KOI-232.04, KOI-232.05, KOI 834.01, KOI 157.03, KOI 707.04 and KOI 1589.03. For KOI 707.04 our estimate is only $1\,m_\oplus$ below the upper limit given by HL17; for KOI-232.05 we might overestimate the planetary density, and the value given by HL17 seems more physically plausible; however, it is significant to less than only $3\,\sigma$. In the other cases we believe that the solutions we provide are more physically plausible than the ones given in HL17.  We refer in more detail to all these specific cases in \S\ref{sec:Individual}.

Overall, there is good agreement between the masses obtained in this work and previously reported planetary masses, thus giving confidence in the fitting process and the reliability of the newly reported masses. The masses, radii, and orbital elements of our adopted solution are tabulated in \S \ref{appendix}, along with machine-readable files.

\clearpage

\begin{figure}[h]
    \includegraphics[width=1\linewidth]{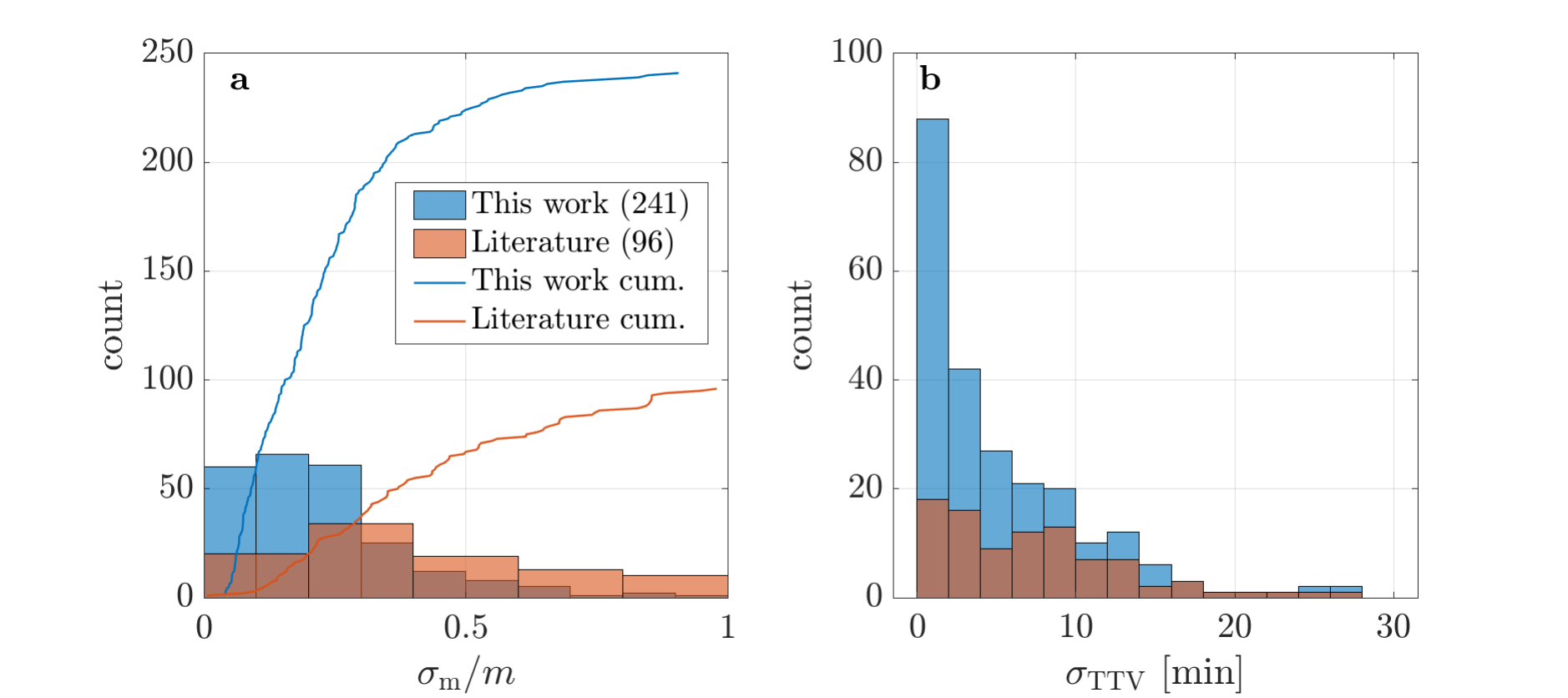}
    \caption{Overall statistics of mass estimates in terms of fractional mass error and TTV magnitude for the planets in this study and our former study (paper II). (a) Distribution of relative error in planetary mass for the combined results of this study and our former study (paper II) (blue) and past literature (orange).  The literature sources taken into account here are detailed in Figure~\ref{fig:MassVsLiterature1}, and in \citet[][, Figure~10]{Judkovsky2022b}. (b) Distribution of $\sigma_{\rm TTV}$, the standard deviation of the TTV signal, in the population of planets for which we obtained a dynamical solution and provided masses in this work and in paper II.\label{fig:MassErrorsHistogram1}}
\end{figure}

In the left panel of Figure~\ref{fig:MassErrorsHistogram1} we show the distribution of fractional error in planetary mass, $\sigma_m/m$, for all the planets for which we obtained a valid dynamical solution in this work (including the 140 planetary masses obtained in paper II). The typical fractional error in planetary mass derived from this study is better than the typical fractional error from past studies by roughly a factor of two. We attribute this improvement to the global fitting approach, which integrates all of the transits together and takes into account all types of transit variations, rather than using TTV fit only (the technique used in most of the referred studies). In addition, on the right panel of the same figure, we show that most of the new planetary masses we obtained are of planets with small TTV magnitudes, quantified by the TTV standard deviation $\sigma_{\rm TTV}$, calculated over the TTV values of all transit events in the best-fitting model.
The current study evidently includes  more TTVs of lower amplitude, and we believe that the explanation to that is that systems with weak dynamical interaction (and, consequently small amplitude TTVs) are usually of small planets, whose individual transit times are anyway poorly constrained. The combination of shallow transits with weak interactions makes it difficult, and sometimes impossible, to extract individual transit times and detect a TTV pattern; in such systems, the global flux fitting can exploit the data better to obtain an estimate of planetary mass.

\clearpage

\subsection{Out-of-Plane Forces}\label{subsec:OutOfPlaneForces}

Forces out of the plane are manifested in transit light curves as Transit Duration Variations (TDVs), or, alternatively, Transit $b$ (impact parameter) Variations (TbVs). In paper II we have shown how TbVs can be interpreted as arising from the interaction among the transiting planets, or from an interaction with a non-transiting companion. 

Only in a handful of cases was mutual inclination constrained from transit variations; however such cases are of high interest and exhibit dynamical richness. A good example is the Kepler-108 system, in which the detection of TDVs led to a photodynamical analysis that showed the planets in this system are highly inclined \citep{MillsFabrycky2017}. Later on, the appearance of such inclination was explained as arising from a binary member via an interaction denoted as  ivection resonance \citep{XuFabrycky2019}. This term refers to a situation in which a resonance between the binary orbital period and the planetary nodal motion acts to pump the planetary orbital inclination. The example of Kepler-108 shows how the detection of a mutually inclined system led to a discovery of a new dynamical phenomenon.

Another avenue in researching mutual inclination in planetary systems involves population analysis of the detected transit variations, such as the usage of the TDV catalog of \citet{Shahaf2021} by \citet{Millholland2021}. The latter authors addressed the distribution of mutual inclinations in planetary systems by analyzing the number of expected TDVs in the case of a dichotomic model that assumes a bi-modal distribution in mutual inclination and a model that assumes that mutual inclinations can be described by a distribution arising from a model based on Angular Momentum Deficit \citep[AMD, ][]{Laskar2017} \citep{He2020}. The results of \citet{Millholland2021} have shown that in terms of the number of detected TDVs, the AMD-based model shows better consistency with the Kepler population than the dichotomic model; this is a significant conclusion, as it offers a non-dichotomic solution to the so-called "Kepler Dichotomy". 
This exemplifies the importance of population analysis of transit variations. It is therefore a natural motivation for us to try expanding the knowledge we have on transit variations arising from forces out of the plane. 

In paper I, we produced a catalog of TbVs of planets in two- and three-transiting planet systems; here, we produce a similar catalog for systems having four or more transiting planets (Table~\ref{tab:SignificantDbdt}). The current catalog contains 71 planets undergoing TbVs to better than $2\,\sigma$, out of which 52 are better than $3\,\sigma$; together with the former catalog from paper II we present 130 planets undergoing TbVs to better than $2\,\sigma$ (77 to better than $3\,\sigma$), an extension to the catalog of Transit Duration Variations (TDVs) published by \citet{Shahaf2021} that contains 31 KOI that display statistically significant TDVs (the only KOI of multiplicity larger than 3 included in their catalog is KOI-841.02, which is also included in our catalog as displaying TbVs).

We note that for most analyzed systems, the signs of the impact parameters are not definitive, despite the typical small errors on $b$. This is because $\lvert b \rvert$ may be well-constrained by the shape of the transit, even if the dynamics do not distinguish between the same-hemisphere configuration  (positive $b$) versus the opposite hemisphere one (in the context of impact parameters and orbital alignment, see also \citet{FabryckyEtAl2014}). We tested this using the following procedure. We examined all of our valid solutions and switched the signs of the impact parameters in all possible configurations. In the language of our fitted parameters, we switched the signs of either $I_x$ or $I_y$ or both for each planet, excluding the innermost planet (for which $I_x$ is set to zero and $I_y$ is limited to positive values only, without the loss of generality). This resulted in $4^{N_{\rm pl}-1}$ configurations for each system, where $N_{\rm pl}$ is the number of planets in the system. For each system, we checked the $\chi^2$ value difference with respect to the original best-fitting solution. Only in two systems do the best-fitting solutions stand out among the others at the level of $2\,\sigma$: the solutions for KOI-834 and for KOI-707. This shows that the signs of $I_x$ and $I_y$ (and with them the sign of $b$) may be inverted in most cases. The two systems mentioned above merit further study, as our solutions constrain their full 3D orbital geometry.

We note an essential difference in the estimation method between our catalog and the catalog of \citet{Shahaf2021}. \citet{Shahaf2021} used individual transit duration measurements and sought significant trends in the duration values for specific KOIs. In other words, they searched for TDVs that could be observed within the time span of the Kepler mission. In this work, we calculate the linear trend in the impact parameter per KOI using the orbital elements' evolution in time; elements that affect all transit variations combined. For many KOIs, we predict a finite TbV trend, although this trend may not be directly seen in the impact parameter measurements. Therefore, our TbV catalog is not a direct empirical observation of TbVs, but a projection of the orbital dynamics on the TbVs. This approach enables predicting the future dynamics of the systems in general: not only discovering faint, slowly varying and periodic TbVs in existing data, but also predicting the detectability of such TbVs not detectable in the time span of the Kepler data.

Having said all that, the results here represent the largest compilation of planets with such variations to date. It is expected that future population analyses, similar to \citet{Millholland2021}, would benefit from this expanded catalog of TbVs.

\begin{figure}[h]
    \includegraphics[width=1\linewidth]{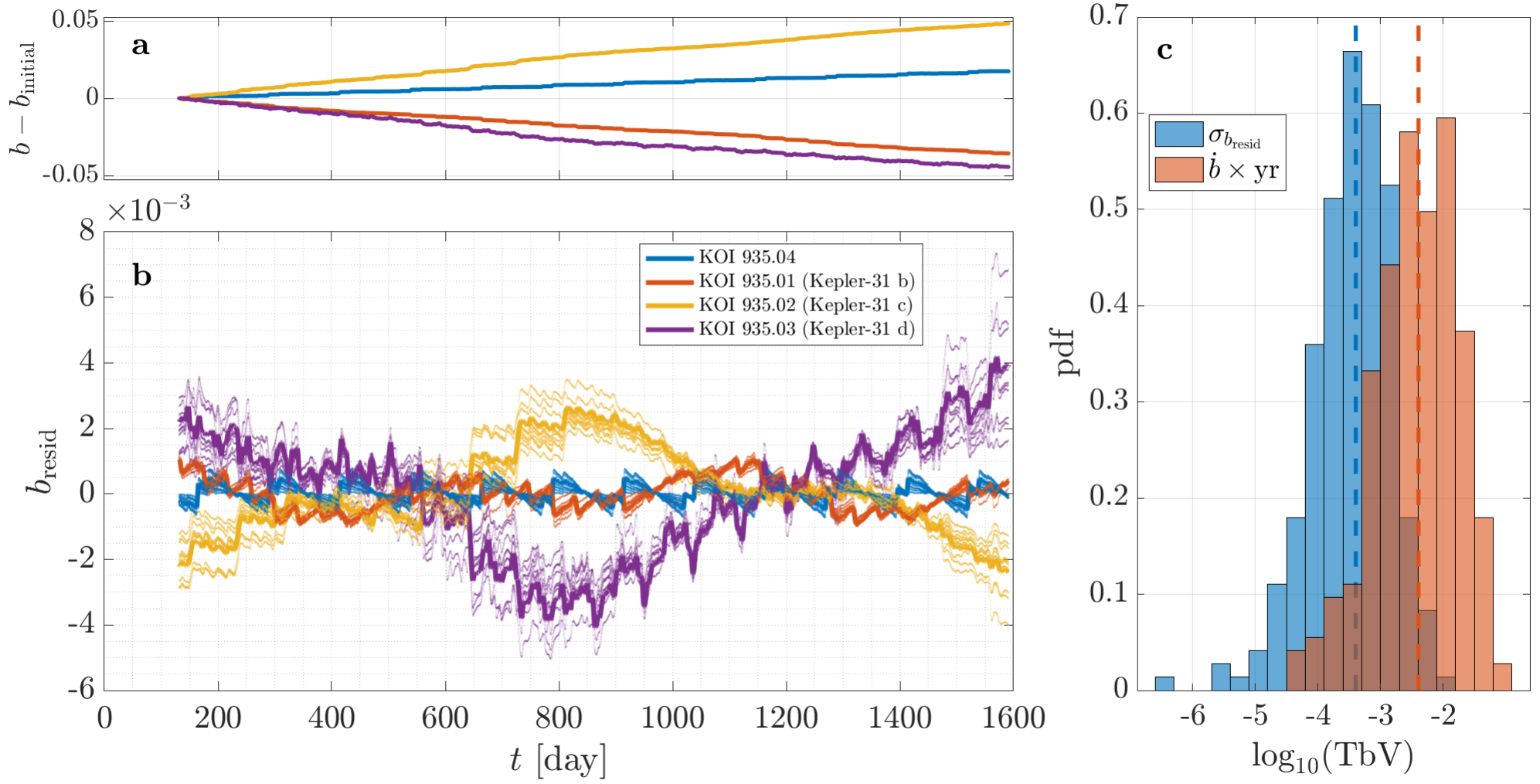}
    \caption{An example for a non-linear TbV pattern, which is superimposed on a secularly-driven TbV linear pattern. (a) The evolution of the impact parameters of the planets in the system KOI-935 (Kepler-31) for the sample median parameter values, shifted by the initial $b$ value, $b_{\rm initial}$, to make the trends visible at the same scale. (b) $b_{\rm resid}$, the residuals of this evolution after the removal of the linear trend (which is attributed to secular interactions). The thick lines show the residuals for the sample median parameter values, and the narrow lines represent the dynamics for ten randomly selected parameter sets from our sample to highlight the non-linear TbV pattern in this system seen in all solutions. The outermost pair, with orbital periods of roughly 42.63 and 87.65 days, reside near but just outside the 2:1 MMR, causing a super-period of roughly 1570 days, which is clearly seen in the non-linear TbV signal. The largest non-linear part of the TbV has a semi-amplitude of $4 \cdot 10^{-3}$ in the best-fitting solution. Such a small signal by itself might not have been detected by direct methods that aim to directly measure the impact parameter for each individual event; it was found here by integrating the information of all transit variations together (timing, duration, depth) in a single dynamical model. The right-hand side panel shows the distribution of $\sigma_{b_{\rm resid}}$ in the best-fitting parameters in the adopted dynamical solutions for 241 planets included in this work and in paper II, and the linearly accumulated TbV over one year for the same sample (using the median value of $\dot{b}$ for each planet). The dashed lines indicate the median values of the two distributions; these differ by a factor of ten. \label{fig:NonLinearDbdt}}
\end{figure}

Incorporating forces out of the plane in the dynamical model has an additional contribution to the masses' determination. The traditional method of using the TTV only can yield mass estimates, but these are sometimes degenerate with eccentricity \citep{LithwickXieWu2012}. Breaking this degeneracy requires either significant eccentricities that would generate a phase-shift between the planets' TTVs, or additional effects, such as the synodic chopping effect\citep{DeckAgol2015} or higher-order TTVs \citep{HaddenLithwick2016}. In Figure~\ref{fig:NonLinearDbdt}, we show an example of a non-linear TbV pattern in addition to the linear secularly-driven TbV drift. We define $b_{\rm resid}$ to be the residual impact parameter variations after the removal of the best-fitting linear trend. The residual TbVs are approximately periodic with a period which is the TTV-derived super-period indicating these are TbV manifestations of the same planet-planet interaction seen in the TTVs. Since \texttt{AnalyticLC} takes into account the various transit variations simultaneously, no additional mechanism is needed to capture this variability. The TbV information adds to the constraining power of the global light-curve approach. On panel c of the same figure, we show the distribution of $\sigma_{b_{\rm resid}}$, which is defined as the standard deviation of $b_{\rm resid}$ over time, of the best-fitting parameters for the 241 planets included in our adopted dynamical solutions in this work and in paper II together. It shows that the typical value of $b_{\rm resid}$ (which is related to near-MMR interactions) is $10^{-4}$ to $10^{-3}$, roughly an order of magnitude smaller than the typical linear change in $b$ (which is related to secular interactions) accumulated in a year. This shows that the light curve contains information content regarding both secular and near-resonant interactions, where the former is more prone to detection for the typical observation time span of a few years. Important for MMR-originating transit variations other than TTV are the TDVs used to extract the properties of the non-transiting planet in the KOI-142 system \citep{NesvornyEtAl2013}.

\begin{figure}[h]
    \includegraphics[width=1\linewidth]{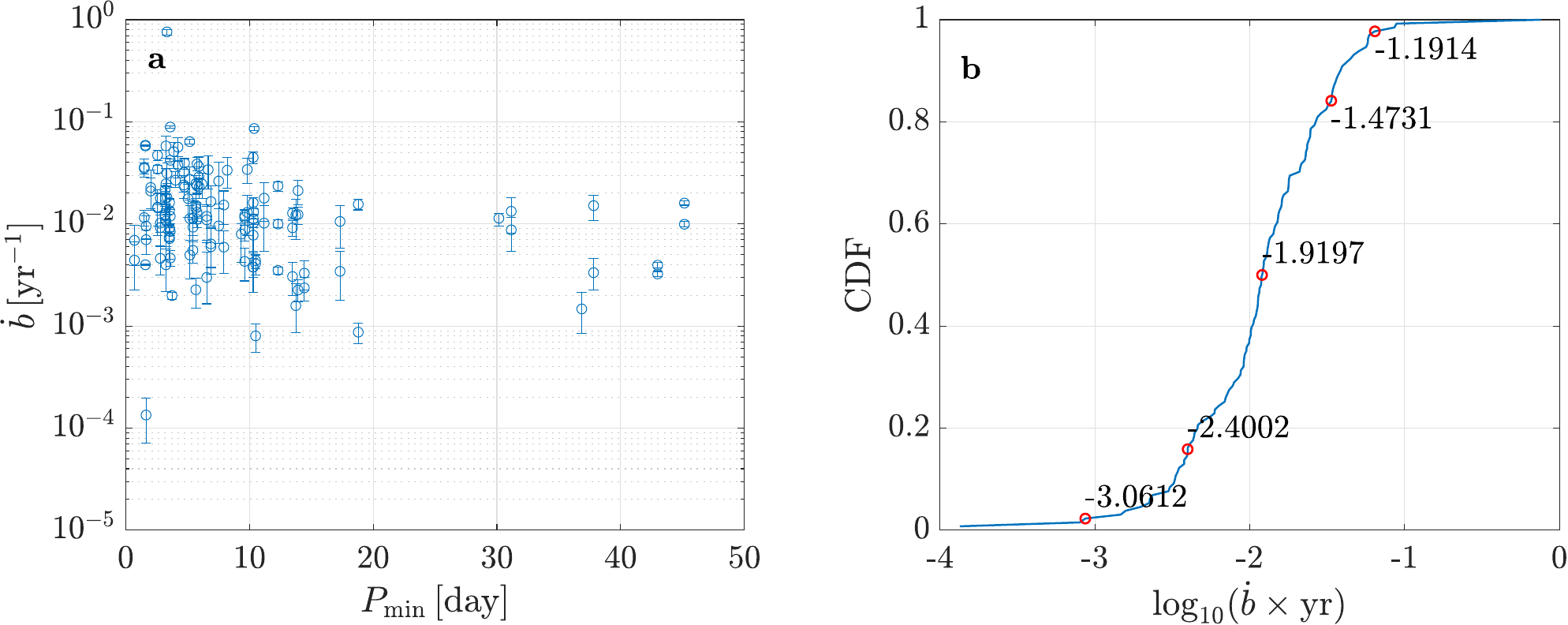}
    \caption{Distribution of TbV magnitude in our 130-planets sample for which $\dot{b}$ is predicted with a significance level larger than $2\,\sigma$, listed in \S\ref{appendix} and in paper II. (a) The absolute value of TbV rate against the shortest-period companion for the 130 planets in our sample that display TbV to better than $2\,\sigma$. (b) the empirical cumulative distribution function of the TbV magnitude in this sample, showing that the typical value is of order $10^{-2}\,\rm yr^{-1}$, with red circles indicating the median and the percentiles equivalent to $1\,\sigma$ and $2\,\sigma$ in a normal distribution..\label{fig:dbdtVsPmin1}}
\end{figure}

We now turn to discussing long-term TbV effects. In Figure~\ref{fig:dbdtVsPmin1}, we show an overview of the 130 values of planets displaying long-term linear TbV with a significance of better than $2\,\sigma$. In the left panel, we show a scatter of these planets' TbV rates against $P_{\rm min}$, the orbital period of the innermost planet in each system, which is a probe of the amount of inward migration the system has experienced. We see that rapid TbV rates occur in systems where the innermost planet's orbital period is of less than 20 days. Specifically, there are eight planets displaying TbV rates faster than $0.05\,\rm yr^{-1}$, seven of which reside in systems that contain planetary companions of less than a five-day period (The exception is of KOI-988.02). 

We note that Figure~\ref{fig:dbdtVsPmin1} shows the dependence of $\dot{b}$ on the \textit{minimal observed orbital period}, not on the period of the planet undergoing the TbV. Some positive correlation between the duration variations rate $\dot{T}$ and the orbital period was proposed by \citet[][their Figure~5]{Shahaf2021}, at least for the significantly constrained $\dot{T}$ values. These are related to our measurements of $\dot{b}$, but are not directly comparable because the transit duration depends on additional factors such as orbital velocity, and the impact parameter depends on geometry alone. 
Note that $\dot{b}$ and $\dot{T}$ arise from different dynamical phenomena: $\dot{b}$ arises mostly due to out-of-plane secular interactions, while $\dot{T}$ can also arise due to in-plane apsidal motion that changes the sky-projected velocity of the planet.

Are systems with short-period planets are expected to display stronger TbVs than other systems? If TbVs are probes of mutual inclination, then there is some supportive evidence for that. \citet{Millholland2020} have shown that the formation of Ultra Short Period (USP) planets, which is a result of inward migration, would require some initial mutual inclination in order to enable the conservation of angular momentum along the migration process. Though we do not focus on USPs here, the same argument implies that systems with short-period planets, which likely migrated to this configuration, would require some initial mutual inclination within the system.
A more quantitative, dynamical analysis of the relation between $\dot{b}$ and $P_{\rm min}$ is left for future work.

\clearpage

To further investigate the population of planets displaying TbV, we define

\begin{equation}
    n_b=\frac{1}{\lvert\dot{b}\rvert}\frac{1}{P}.
\end{equation}

\begin{figure}[h]
    \includegraphics[width=1\linewidth]{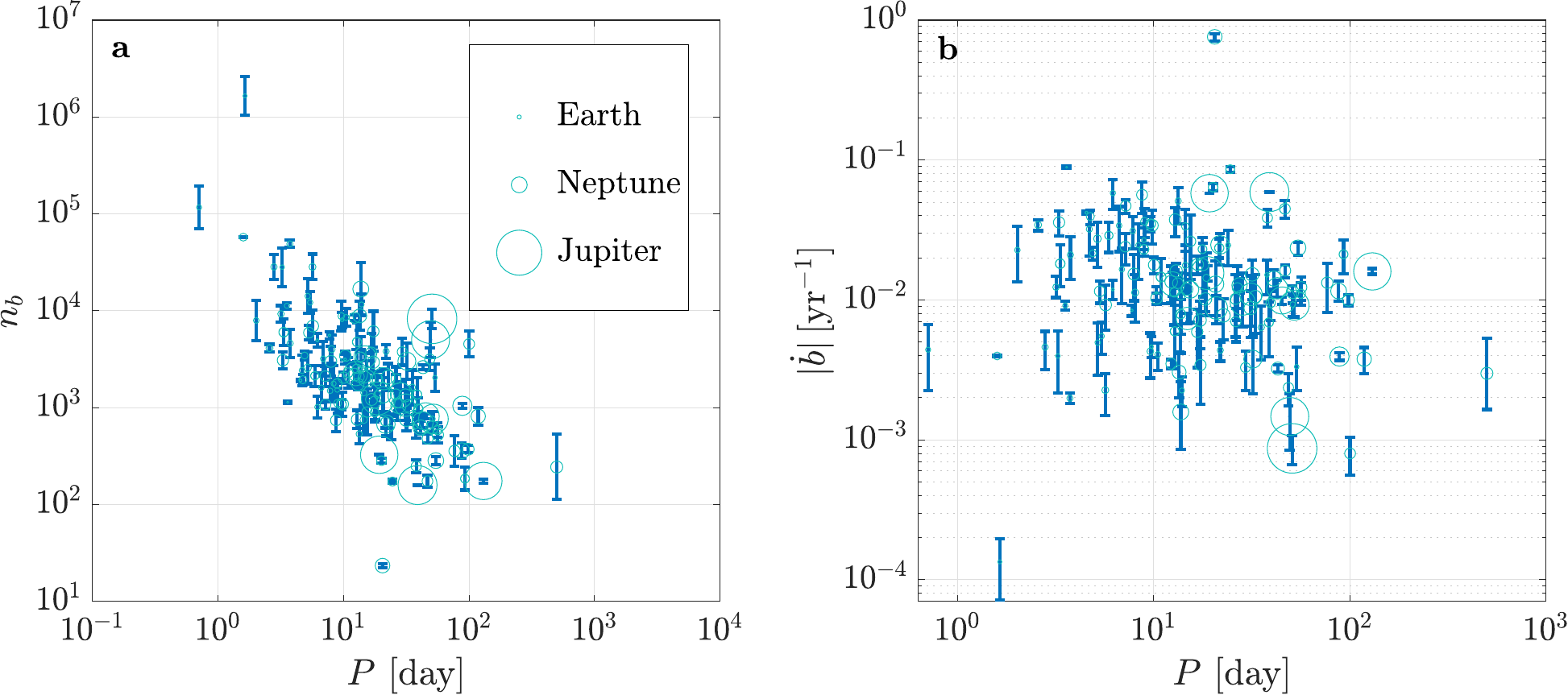}
   \caption{Measured impact parameter variation rates. (a) Timescale for transits to disappear (namely, number of orbital periods for $b$ to change by 1) against an orbital period of 130 Tbv-displaying planets. The radii of the circles represent the absolute planetary radii. (b) Absolute values of $\dot{b}$ versus orbital period. The values of $\dot{b}$ are spread around $\approx 10^{-2}$ across the range of observed orbital periods. The correlation of $\lvert\dot{b}\rvert$  with $P$ is weak. 
    \label{fig:dbdtTimeScale1}}
\end{figure}

The dimensionless quantity $n_b$ is the characteristic number of orbits required for the planet to change its impact parameter by unity (given a constant change rate), i.e. a currently transiting planet will be torqued out-of transit after typically $n_b$ transits. In Figure~\ref{fig:dbdtTimeScale1} we plot this number against the orbital period of the TbV-displaying planets. We find that $\log{n_b}$ varies roughly inversely with $P$, that is that the measured $\dot{b}$ values are only weakly correlated with the orbital period (panel b). If this finding holds also for larger periods (which is not known), then planets of orbital periods of decades would typically be torqued out of transit after just a few orbital periods. This would imply that long-period planets not only have a lower geometric transit probability \cite{RagozzineHolman2010} but also that they tend to move out of transit in a small number of orbits, limiting the utility of the transit method. Further study is required in order to shed light on this point. We resist fitting the $n_b(P)$ trend because we find the fit parameters sensitive to the removal of a small number of points from the sample.

\begin{figure}
    \includegraphics[width=1\linewidth]{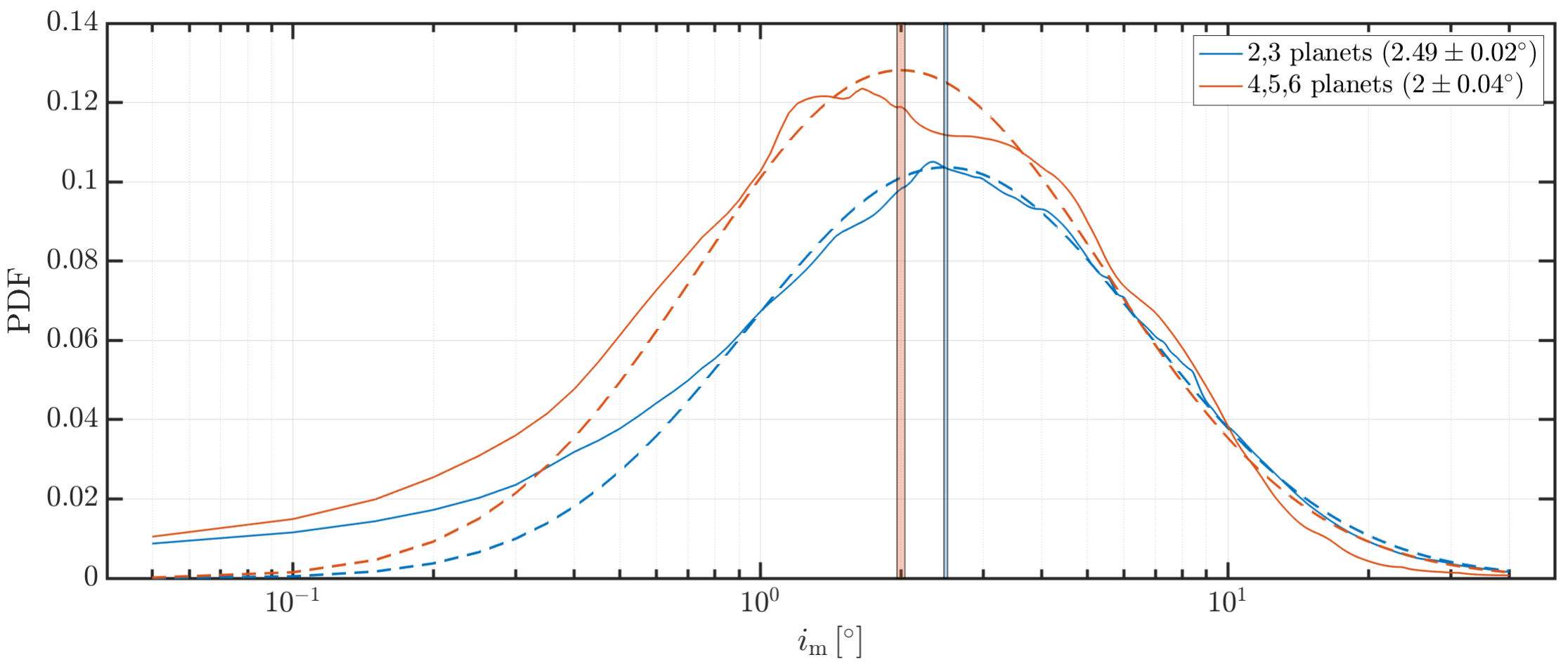}
    \caption{Distribution of mutual inclinations among all pairs in all our runs that passed the dynamical validity criterion $\sigma_{\rm Nbody}<1.5$, and after shuffling over all configurations of this signs of $I_x$ and $I_y$. Each solid curve is obtained by averaging all empirical PDFs from all the runs passing the criterion above calculated over uniform bins in $i_m$ and shown in a logarithmic scale. The dashed lines indicate the best-fitting Gaussian for each curve: blue for systems with 2 or 3 planets and red for systems with more than 3 planets. The vertical lines indicate the 99\% confidence intervals of the expectation values of the fits (also listed in the legend) which are $2.49\pm 0.02^{\circ}$ for systems with 2,3 planets and  $2\pm 0.04^{\circ}$ for systems with more than 3 planets. This gives a qualitatively similar picture to that of \citet{He2020} in both the distribution shape and its dependence on multiplicity.\label{fig:MutIncDistribution}}   
\end{figure}

In addition to the TbV analysis, we also examine the direct output of the dynamical fitting: the mutual inclinations among the planets. In Figure~\ref{fig:MutIncDistribution} we show the empirical distribution of mutual inclinations obtained from all our runs that passed all our validity tests, including a summation of all configurations obtained after changing the signs of $I_x$ and $I_y$ in systems whose fits permitted this inversion (as explained above). This figure shows that the mutual inclination among the transiting planets, $i_m$, distributes in a log-normal shape, and indicates that systems with more transiting planets tend to have somewhat smaller mutual inclinations. Both the shape of the distributions and the trend of multiplicity against mutual inclination match the findings of \citet[][their Figure~6]{He2020}, who proposed that the statistics of planetary systems properties can be explained by an AMD \citep[angular momentum deficit,][]{Laskar2017} based model. Their proposed model was later supported by \citet{Millholland2021}, based on the statistics of transit duration \citep{Shahaf2021}.
We note that in many systems, both in this work and in paper II, we propose the existence of a non-transiting external perturber. This means that the inclination values obtained in those systems may be suspect. Nonetheless, the qualitative similarity to the results presented by \citet[][their Figure~6]{He2020} is compelling because their method relied on detection statistics of the Kepler population, while ours relies on dynamical modeling in individual systems. The fact that both methods yield similar findings supports the proposed AMD-based model.

\clearpage

\section{Individual Systems Description}\label{sec:Individual}

\begin{figure}[h]
    \includegraphics[width=1\linewidth]{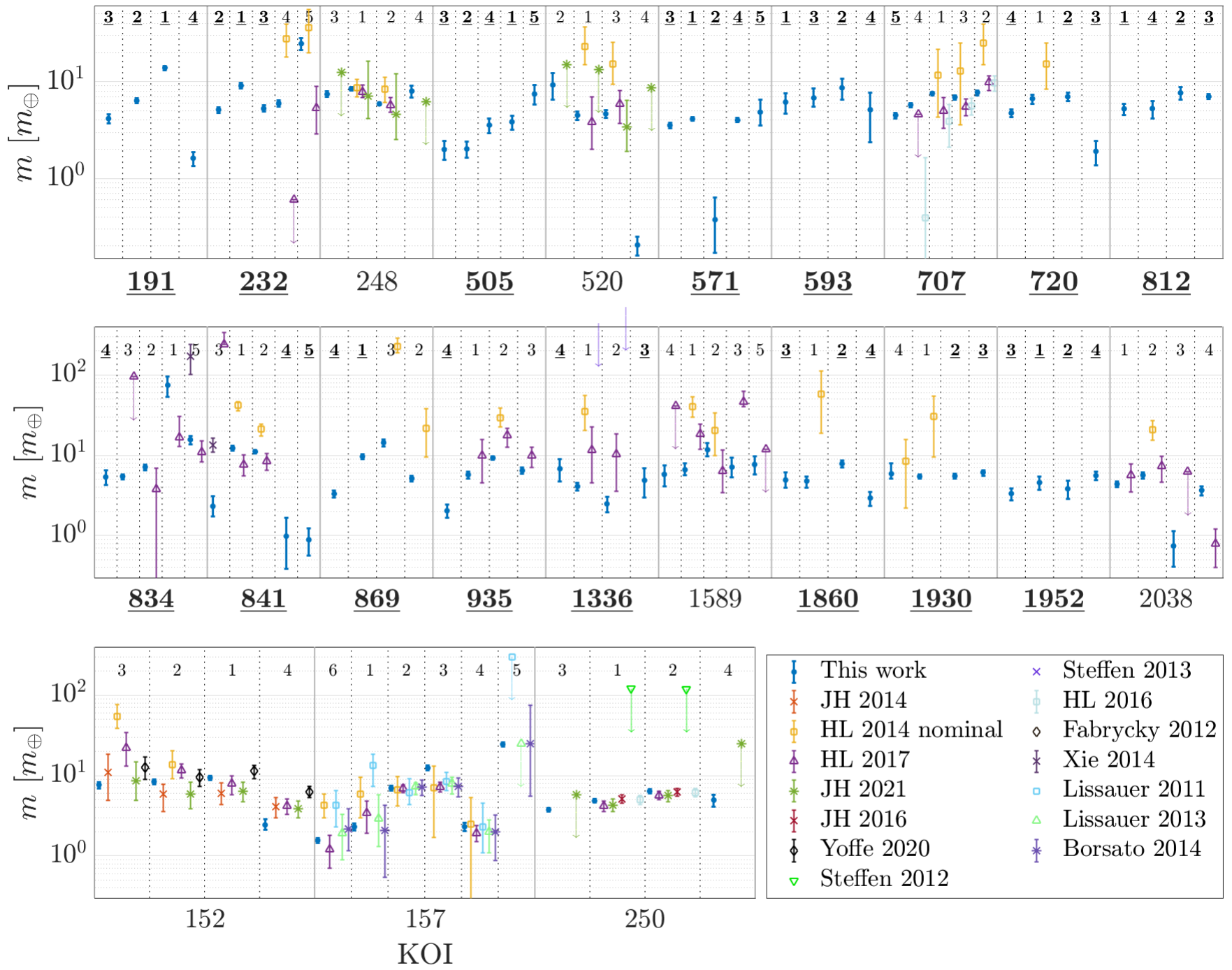}
    \caption{Comparison of the masses obtained from this work with past literature by KOI. In each subpanel, the planets in the system are sorted (left to right) by their orbital period. The numbers above are the suffix of the planets' KOI designation; for instance, the first object in the plot is KOI-191.03. Dotted vertical lines separate different planets within a system. Bold underlined KOI numbers are used for systems with at least one planet without a previously reported mass constraint, as in Figure~\ref{fig:FittedPlanetsOverview2}. The bold underlined KOI suffix represents a specific planet without a previously reported mass. The legend shows the colors and markers representing different literature sources. Arrows pointing down represent upper limits on mass.
    \label{fig:MassVsLiterature1}
    }
\end{figure}

In this section, we provide details for each of the 23 systems with a dynamical solution. We review previous literature on planetary masses as well as highlight the dynamical features of the system. The systems are ordered by their KOI index.
The comparison to former literature masses is given in Figure~\ref{fig:MassVsLiterature1} for all the KOIs. We include both past estimates of the true planetary mass and estimates of planetary nominal mass \citep[][hereon HL14 and X14]{HaddenLithwick2014,Xie2014}, which is, in many cases, an upper limit on the true mass. For the values of \citet[][hereon HL17]{HaddenLithwick2017}, we use their high-mass prior (which is appropriate for comparing with our results, which are obtained from using the same prior as theirs).

From our sample of \Nmasses planets, for \Nnewmasses we did not find any previous literature mass value. For \NnewmassesWithThreeSigma out of those, our mass detection is significant to more than $3\,\sigma$.
For ten planets, we obtained a median mass estimate lower than $2\,m_\oplus$, of which six are lower than $1\,m_\oplus$: KOI numbers 248.04, 520.04, 571.02, 841.04, 841.05, 2038.03. For three of those (248.04, 520.04, 2038.03) we found the results consistent with the upper limits given by HL17 and JH21, for the other three (841.04, 841.05, 571.02), we did not find any previous mass constraints.
Out of \Nmasses planets in systems with a valid solution, for \NmassesWithThreeSigma, we provide mass constraints better than $3\,\sigma$.

\ksec{Kepler-79 (KOI-152)}
All our runs converged to solutions of similar masses and to nearly circular orbits. One of the runs yielded a solution significantly better than all others regarding fit quality; it indicates small mutual inclinations of a few degrees. The outermost transiting planet (KOI-152.04, Kepler-79 e) is almost grazing, with an (absolute value of the) impact parameter of roughly 0.96 and a planet-to-star radius ratio of about 0.026. The two intermediate planets (KOI-152.02 and KOI-152.01, namely Kepler-79 c and Kepler-79 d) display significant TbVs of order $0.01\,{\rm yr}^{-1}$, and the innermost planet (b) displays significant, somewhat slower, TbVs of order $0.003\,{\rm yr}^{-1}$. The best model TTV nicely matches the TTV data of \citet{HolczerEtAl2016}.
Our adopted solution suggests masses of $7.7^{+0.6}_{-0.7}\,m_\oplus$ for KOI 152.03 (Kepler-79 b), $8.4^{+0.5}_{-0.7}\,m_\oplus$ for KOI 152.02 (Kepler-79 c), $9.4^{+0.6}_{-0.6}\,m_\oplus$ for KOI 152.01 (Kepler-79 d), $2.4^{+0.4}_{-0.3}\,m_\oplus$ for KOI 152.04 (Kepler-79 e). These are within $\sim1.5\,\sigma$ of the results of \citet{JontoffHutter2014}, within $\sim1.5\,\sigma$ of the results of HL17 in their high-mass prior case, and consistent with the results of JH21. 
The solution suggests a clear distinction between the densities of the two inner planets ($\sim1-1.5\,{\rm g\,cm^{-3}}$) and the two outer planets ($\sim0.15,0.25\,{\rm g\,cm^{-3}}$).

\ksec{Kepler-11 (KOI-157)}
This system, which contains six transiting planets, has been examined in many  studies. \citet{Lissauer2011a} analyzed the TTVs known at that time to estimate  the planetary masses and studied the system's dynamical stability. These masses were then estimated again \citep{Lissauer2013} using more (14 quarters) of Kepler data, and later again by \citet{Borsato2014}. HL14 estimated the nominal masses and, later (HL17), the true masses of the planets in this system. Our results agree well with these studies, apart from the fourth known planet from the star (Kepler-11 e, KOI 157.03), where our adopted solution suggests mass of $12.5^{+0.8}_{-0.9}\,m_\oplus$ while \citet{Lissauer2013} suggests $8^{+1.5}_{-2.1}\,m_\oplus$ and HL17 suggest $7.2^{+1.1}_{-1.0}\,m_\oplus$. 
The period ratios in this system suggest that each planet TTV is affected by more than one specific frequency; this, along with the relatively massive planets and the wealth of information encoded in six transit signals, make the solution distinct with multiple studies in agreement.

\ksec{Kepler-487 (KOI-191)}
We found no literature mass for the four KOIs in this system. Until recently, the inner two were designated as candidates, and the outer two as confirmed. Currently, KOI 191.02 (planet Kepler-487 d) is designated as confirmed in NExSci following the analysis of \citet{Valizadegan2021}.

This system is of dynamical interest for several reasons. First, of the four transiting planets it contains, the third one (planet b, orbital period of 15.35 days), displays a transit depth that suggests a Jupiter-size object, not a frequent scenario in compact Kepler systems. Second, this system harbors an Ultra Short Period (USP) planet \citep{Winn2018}. This type of planet has probably gone through a runaway migration process which requires strong dissipation mechanisms \citep[for example ][and references therein]{Millholland2020}. Third, unlike many Kepler compact systems, the planets here are not near any resonance.

We provide one solution, with masses of $4.1^{+0.4}_{-0.4}\,m_\oplus$ for KOI 191.03, $6.3^{+0.3}_{-0.4}\,m_\oplus$ for KOI 191.02, $13.9^{+0.7}_{-0.8}\,m_\oplus$ for KOI 191.01 (Kepler-487 b), $1.6^{+0.2}_{-0.3}\,m_\oplus$ for KOI 191.04 (Kepler-487 c). We note that this solution involves a large eccentricity of the innermost 0.7-day orbit planet. Such a situation would be surprising, as short-orbit planets should have their orbits circularized on short timescales. 

A possible explanation we suggest is that though our pipeline converged to a four-planet model that fits the data, the system contains more than four planets. A non-transiting fifth planet could possibly explain the high-amplitude TTV ($\sim13.4$ minutes) of the outermost transiting planet (planet c, KOI 191.04), without the need for high-eccentricity orbits.

We, therefore, provide the four-planet solution of the system but note that it may require further investigation in order to understand its dynamical structure. 

\ksec{Kepler-122 (KOI-232)}
We provide one solution, that suggests masses of $5.1^{+0.3}_{-0.4}\,m_\oplus$ for KOI 232.02, $9.1^{+0.6}_{-0.6}\,m_\oplus$ for KOI 232.01, $5.3^{+0.4}_{-0.4}\,m_\oplus$ for KOI 232.03, $6^{+0.4}_{-0.4}\,m_\oplus$ for KOI 232.04, $24.7^{+3.7}_{-3.5}\,m_\oplus$ for KOI 232.05. HL17 analyzed the interaction between planet e (KOI 232.04) and planet f (KOI 232.05), which display strong TTVs of order 45 and 10 minutes, respectively.

For KOI-232.04 (Kepler-122 e ), HL17 give an upper mass limit of $0.6\,m_\oplus$ based on their high-mass prior.
Given the radius of order $2.5\,R_\oplus$, our solution is more physically plausible. For KOI-232.05 (Kepler-122 f) HL17 give $5.3^{+3.6}_{-2.4}\,m_\oplus$, a mass significantly lower than ours. With a radius of $\sim2.3\,R_\oplus$, our solution implies an abnormally high density. 

An alternative explanation for the possible overestimate of Kepler-122 f density is that this system harbors another external non-transiting companion. Such a scenario could explain the TTVs of Kepler-122 e without the need for a high mass for Kepler-122 f, and at the same time explain the significant TbVs displayed by the planets in this system (see Table \ref{tab:SignificantDbdt}) without the need for the few-degrees mutual inclinations suggested by the dynamical solution (see Table \ref{tab:PhysicalOrbital}).

\ksec{Kepler-49 (KOI-248)}
HL17 analyzed the interaction between KOI-248.01 and KOI-248.02 (Kepler-49 b and Kepler-49 c, respectively). Taking the joint errors, our solution agrees with HL17 values and differs from the values provided by \citet{JontofHutter2016} by $\sim2\,\sigma$. 
A surprising result of our best-fitting solution is the implied mutual inclination among the planets. The solution suggests a mutual inclination of the innermost planet from the other planets of $\sim25^\circ$. 
This tilt is unexpectedly high but may have a real dynamical origin. 
This innermost planet is at a period ratio of 2.8 with the second planet, not close to any first-order Mean Motion Resonance (MMR), while the second and third planets are close to the 3:2 MMR. The short orbital period of the innermost planet (2.58 days) approaches the one-day boundary of the orbits of USP planets \citep{Winn2018}. Such planets are thought to have undergone rapid inward migration, during which angular momentum conservation requires that they possess initial orbital inclination \citep{Millholland2020}. We highlight this system, as it might include a planet in the process of becoming a USP. Future observations of the orbital period of the innermost planet could reveal if this is the case.

\ksec{Kepler-26 (KOI-250)}
The planets in this system were confirmed based on \citet{Steffen2012}, who analyzed their TTVs and have shown that the TTVs are anti-correlated and hence the planets are dynamically interacting.
Our solution suggests masses of $3.8^{+0.2}_{-0.2}$ for KOI 250.03 (Kepler-26 d), $4.8^{+0.3}_{-0.3}$ for KOI 250.01 (Kepler-26 b), $6.4^{+0.4}_{-0.3}$ for KOI 250.02 (Kepler-26 c), $5^{+0.8}_{-0.9}$ for KOI 250.04 (Kepler-26 e). The masses of Kepler-26 b and Kepler-26 c agree with those given by \citet{HaddenLithwick2016}, \citet{JontofHutter2016} and HL17. The masses of Kepler-26 d and Kepler-26 e agree with the upper limits given by JH21.
Our solution involves eccentricity of order 0.1 for the innermost 3.54 days planet and inclinations of about 10 degrees; these are relatively high values, and we suggest that further observations would be needed in order to confirm them.

\ksec{Kepler-169 (KOI-505)}
We found no literature values for the masses in this five-transiting planet system.
We provide two solutions; although one of them is of better-fit quality by $\sim2\,\sigma$, we select the other one as the adopted solution due to the more plausible density it suggests for the innermost planet.
Both solutions suggest small eccentricities of a few percent and inclinations of a few degrees, which are also expressed as $2\,\sigma$ TbV signals of the four innermost planets.
The adopted solution suggests a monotonic mass order, from $\sim1\,m_\oplus$ of the innermost planet to $\sim7\,m_\oplus$ for the outermost planet.

\ksec{Kepler-176 (KOI-520)}
All adjacent pairs in this four-transiting planet system are near the 2:1 MMR.
HL17 estimated the masses of KOI-520.01 (Kepler-176 c) and KOI-520.03 (Kepler-176 d): $3.8^{+3.2}_{-1.8}\,m_\oplus, 5.9^{+2.2}_{-2.8}\,m_\oplus$, respectively. JH21 provided upper limits consistent with these values.
Our solution suggests masses of $9.2^{+3}_{-2.7}\,m_\oplus$ for KOI 520.02 (Kepler-176 b), $4.5^{+0.4}_{-0.4}\,m_\oplus$ for KOI 520.01, $4.6^{+0.4}_{-0.4}\,m_\oplus$ for KOI 520.03, $0.2^{+0.05}_{-0.05}\,m_\oplus$ for KOI 520.04 (Kepler-176 e), consistent with the results of HL17. The outer planet's mass and density are exceptional; its derived mass is one of the lowest known of any exoplanet, and with a radius of about $1.36\,R_\oplus$, it possesses a density of $0.44^{+0.11}_{-0.11}\,{\rm g\,cm^{-3}}$, a surprisingly low value for a small planet below the radius gap. The eccentricities are a few percent.
The innermost and outermost planets display significant TbVs, explained in this solution by mutual inclinations of 8-9 degrees.
Based on these unusual characteristics, we highlight this system for further observations, and suspect that there is an additional planet(s) in the system that affects the observed dynamics. Such a planet might explain the low-mass and eccentric orbit of the outermost planet, KOI-520.04, suggested by our obtained solution.

\ksec{Kepler-186 (KOI-571)}
We found no literature values for the masses in this five-transiting planet system.
We provide one solution with significant mass constraints of $>3\,\sigma$ for four out of the five planets, with masses of $3.5^{+0.2}_{-0.2}\,m_\oplus$ for KOI 571.03, $4.1^{+0.2}_{-0.2}\,m_\oplus$ for KOI 571.01, $0.4^{+0.3}_{-0.2}\,m_\oplus$ for KOI 571.02, $4^{+0.2}_{-0.2}\,m_\oplus$ for KOI 571.04, $4.8^{+1.6}_{-1.3}\,m_\oplus$ for KOI 571.05  The mass estimate of KOI-571.02 (Kepler-186 d) is surprisingly low for a $1.52\,R_\oplus$ planet, as it is difficult to retain a substantial atmosphere with such a low mass. In addition, the solution includes a high eccentricity value of 0.2 for the innermost, 3.88-day orbit planet, KOI-571.03 (Kepler-186 b). We hence consider this solution as questionable and conclude that further observations would be required to constrain the parameters of this system.

\ksec{Kepler-616 (KOI-593)}
We found no literature mass for this four-planet system.
We provide two solutions, which are consistent in most parameters and differ only in some of the eccentricity components. Our adopted solution suggests masses of $6.1^{+1.5}_{-1.5}\,m_\oplus$ for KOI 593.01 (Kepler-616 b), $6.8^{+1.7}_{-1.3}\,m_\oplus$ for KOI 593.03, $8.6^{+2.2}_{-2.1}\,m_\oplus$ for KOI 593.02 (Kepler-616 c), $5.1^{+2.6}_{-2.8}\,m_\oplus$ for KOI 593.04; the three innermost transiting planets have a significant mass detection of more than $3\,\sigma$. The solution suggests eccentricities of a few percent and inclinations of a few degrees.

\ksec{Kepler-33 (KOI-707)}
This five-planet system has been analyzed by HL14, \citet{HaddenLithwick2016}, and HL17.
Our solution suggests masses of $4.5^{+0.3}_{-0.4}\,m_\oplus$ for KOI 707.05 (Kepler-33 b), $5.7^{+0.3}_{-0.3}\,m_\oplus$ for KOI 707.04 (Kepler-33 c), $7.6^{+0.3}_{-0.4}\,m_\oplus$ for KOI 707.01 (Kepler-33 d), $6.9^{+0.3}_{-0.4}\,m_\oplus$ for KOI 707.03 (Kepler-33 e), $7.7^{+0.4}_{-0.5}\,m_\oplus$ for KOI 707.02 (Kepler-33 f); those masses differ from the results of HL17 by 1.5-2 $\sigma$. This difference might partially arise from the TbVs of the outermost planet, which is not taken into account in a TTV-only analysis. 
Our solution suggests small eccentricities of a few percent but significant inclinations of up to 10 degrees; an alternative explanation for the observed TbVs in this system could be an external, non-transiting companion.
This system contains a fingerprint of the SMMR effect we mentioned in paper I: planets Kepler-33 d and Kepler-33 e are slightly inside the 3:2 MMR, with a super-period of roughly 393.76 days, and planets Kepler-33 e and Kepler-33 f are slightly inside the 4:3 MMR with a super-period of approximately 321.7 days. The relative proximity of these super-periods yields a TTV pattern arising from the SMMR effect, for which we estimate an amplitude of a few minutes at a timescale of roughly 4.81 years.
In addition, this system is one of two (along with KOI-834) in which inverting the impact parameter signs of the best-fitting solutions yields a fit quality worse by $\sim\,2\,\sigma$ (see \S~\ref{subsec:OutOfPlaneForces}). This gives us confidence that the impact parameter signs are meaningful, and therefore these two systems deserve further investigation regarding their 3D structure.

\ksec{Kepler-221 (KOI-720)}
For this system of four transiting planets, we found in the literature a nominal mass for planet c (HL14).
Our solution suggests masses of $4.7^{+0.4}_{-0.4}\,m_\oplus$ for KOI 720.04 (Kepler-221 b), $6.7^{+0.6}_{-0.7}\,m_\oplus$ for KOI 720.01 (Kepler-221 c), $7^{+0.7}_{-0.7}\,m_\oplus$ for KOI 720.02 (Kepler-221 d), $1.9^{+0.5}_{-0.6}\,m_\oplus$ for KOI 720.03 (Kepler-221 e), where the innermost planet is of a density slightly higher than Earth and the other planets with densities lower than Earth. 
The two outermost planets display significant TbVs; these may relate to the 5-10 degrees of mutual inclinations suggested by the solution; an alternative explanation could be an external non-transiting companion.

\ksec{Kepler-235 (KOI-812)}
We found no literature mass for this four-planet system. Our solution suggests masses of $5.2^{+0.7}_{-0.7}\,m_\oplus$ for KOI 812.01 (Kepler-235 b), $5.3^{+1.1}_{-1.1}\,m_\oplus$ for KOI 812.04 (Kepler-235 c), $7.6^{+1.1}_{-1.2}\,m_\oplus$ for KOI 812.02 (Kepler-235 d), $7^{+0.4}_{-0.4}\,m_\oplus$ for KOI 812.03 (Kepler-235 e). The solution suggests eccentricities of up to 0.1 and inclinations as high as 10 degrees (though with large error bars). In addition, the solution suggests that the innermost transiting planet KOI-812.01, which is above the radius gap, has a similar mass to the second planet (KOI-812.04); it is a surprising result because with similar masses it would be more likely for KOI-812.01 to lose its atmosphere than for KOI-812.04. Given all of the characteristics of the solutions above, we conclude that more data would be required to constrain the planetary masses in this system. It's entirely possible that there is a non-transiting companion in this system, and by not taking it into account, we overestimated the mass of KOI 812.04.

\ksec{Kepler-238 (KOI-834)}
HL17 analyzed the interaction among the four outer planets out of the five transiting planets. These four planets (c,d,e,f) form a near-resonant 2:1 chain. 
We provide different solutions, suggesting masses of $5-10\,m_\oplus$ for the three innermost planets, a giant-planet mass of order $70\,m_\oplus$ for the fourth planet, and mass of $10-15\,m_\oplus$ for the outermost planet. All solutions suggest small eccentricities of a few percent. Most of them, including the adopted one, indicate inclinations of a few degrees.
The high-prior mass values of HL17 for planet c are insignificant; our results for planets d and f are consistent with HL17 up to the uncertainty. Our solution for planet e suggests a higher mass than in the solution of HL17; we find our solution physically plausible as the planetary radius is about $9\,R_\oplus$. 
This system is one of two (along with KOI-707) in which inverting the impact parameter signs of the best-fitting solutions yields a fit quality worse by $\sim\,2\,\sigma$ (see \S~\ref{subsec:OutOfPlaneForces}). This gives us confidence that the impact parameter signs are meaningful, and therefore these two systems deserve further investigation regarding their 3D structure.

\ksec{Kepler-27 (KOI-841)}
Two planets are confirmed in this five-transiting KOIs system (KOI-841.01 is Kepler-27 b, KOI-841.02 is Kepler-27 c), while the others are considered candidates. HL17 estimated the masses of planets b and c and the mass of the innermost KOI-841.03.
We provide a solution that suggests masses of $2.3^{+0.8}_{-0.6}\,m_\oplus$ for KOI 841.03, $12.2^{+0.9}_{-0.9}\,m_\oplus$ for KOI 841.01, $11.1^{+0.6}_{-0.6}\,m_\oplus$ for KOI 841.02, $1^{+0.7}_{-0.6}\,m_\oplus$ for KOI 841.04, $0.9^{+0.4}_{-0.3}\,m_\oplus$ for KOI 841.05.
The two outermost planets, which are of long periods, are not significantly constrained. The suggested masses of planets b and c are slightly higher than the values given by HL17. The indicated eccentricity values of the innermost planet, of order 0.1, are high for a 6.54-day orbit planet; hence, this dynamical solution is suspicious. An alternative explanation could be the existence of non-transiting companions in the system, which could also explain the TbV signals of KOI-841.01 (Kepler-27 b), KOI-841.02 (Kepler-27 c), and KOI-841.05 without resorting for the significant inclination of KOI-841.05 suggested by the solution.

\ksec{Kepler-245 (KOI-869)}
We found in the literature nominal masses for the two outermost planets given by HL14. 
We provide three dynamical solutions with similar fit quality. The adopted one is selected based on its coplanar structure, likely in a four-transiting-planets system \citep{RagozzineHolman2010}. Our adopted solution suggests masses of $3.3^{+0.3}_{-0.4}\,m_\oplus$ for KOI 869.04 (Kepler-245 e), $9.7^{+0.6}_{-0.7}\,m_\oplus$ for KOI 869.01 (Kepler-245 b), $14.4^{+1.4}_{-1.5}\,m_\oplus$ for KOI 869.03 (Kepler-245 c), $5.1^{+0.4}_{-0.4}\,m_\oplus$ for KOI 869.02 (Kepler-245 d). 
No significant TbVs are observed.

\ksec{Kepler-31 (KOI-935)}
This four-transiting-planet system has been analyzed by \citet{Fabrycky2012}, who used TTVs from the first eight Kepler quarters to conclude that the transit signals arising from KOI-935.01 (Kepler-31 b) and KOI-935.02 (Kepler-31 c) arise from the same host star. HL17 estimated the masses of KOI-935.01 (Kepler-31 b), KOI-935.01 (Kepler-31 c), KOI-935.01 (Kepler-31 d); KOI 935.04 is currently designated as a candidate.
We provide three solutions, and select the adopted one due to fit quality; it is better than the others by $\Delta\chi^2\sim230$, equivalent to $\sim7\,\sigma$.
Our adopted solutions suggests masses of $2^{+0.4}_{-0.4}\,m_\oplus$ for KOI 935.04, $5.7^{+0.6}_{-0.7}\,m_\oplus$ for KOI 935.01, $9.3^{+0.4}_{-0.5}\,m_\oplus$ for KOI 935.02, $6.5^{+0.5}_{-0.6}\,m_\oplus$ for KOI 935.03. All of these masses are within 1.5 sigma of the HL17 masses. All planets display significant impact parameter variations; this is probably the source of the relatively high mutual inclinations suggested by the solutions. A possible alternative explanation for the impact parameter variations would be an external non-transiting companion that applies torques on the transiting planets.

\ksec{Kepler-58 (KOI-1336)}
\citet{Steffen2013} have shown that the TTVs of KOI-1336.01 (Kepler-58 b) and KOI-1336.02 (Kepler-58 c) are anti-correlated and hence reside in the same system, based on the data available at the time. Later, HL14 estimated the nominal masses of KOI-1336.01 (Kepler-58 b) based on the TTV amplitude of KOI-1336.02 (Kepler-58 c). HL17 provided masses for those planets with large error bars. 
We provide three solutions, all with similar fit quality. The adopted solution is selected based on best matching with $N$-body integration. Our adopted solution suggests masses of $6.8^{+2.2}_{-2.1}\,m_\oplus$ for KOI 1336.04, $4.1^{+0.4}_{-0.5}\,m_\oplus$ for KOI 1336.01, $2.5^{+0.6}_{-0.5}\,m_\oplus$ for KOI 1336.02, $4.9^{+2}_{-1.9}\,m_\oplus$ for KOI 1336.03 (Kepler-58 d).

\ksec{Kepler-84 (KOI-1589)}
This five-planet system has been analyzed by HL14 for nominal masses, and later by HL17.
We provide two solutions; the adopted one is slightly better regarding fit quality. The solutions agree on the planetary masses, apart from slightly different values for KOI-1589.03 (Kepler-84 e). The adopted solution suggests masses of $5.8^{+1.7}_{-1.7}\,m_\oplus$ for KOI-1589.04 (Kepler-84 d), $6.6^{+1.3}_{-0.9}\,m_\oplus$ for KOI-1589.01 (Kepler-84 b), $11.7^{+2.5}_{-2}\,m_\oplus$ for KOI-1589.02 (Kepler-84 c), $7.1^{+2.2}_{-1.8}\,m_\oplus$ for KOI-1589.03 (Kepler-84 e), $7.7^{+2.1}_{-1.9}\,m_\oplus$ for KOI-1589.05 (Kepler-84 f).
Our solution agrees with the upper limits given by HL17 on KOI-1589.04 and KOI-1589.05 and with the values given for planet c. 
Our values are smaller than those given by HL17 for KOI-1589.01 and KOI-1589.03.
KOI-1589.02 displays a significant TbV signal ($\sim4\,\sigma$).
This system displays a small-amplitude TTV pattern due to the SMMR effect we mentioned in paper I: Kepler-84 b and Kepler-84 c are slightly inside the 3:2 MMR with a super-period of roughly 273.12 days, and Kepler-84 c and Kepler-84 e are outside the 2:1 MMR with a super-period of approximately 211.84 days. The relative proximity between these super-periods yields a TTV pattern at a timescale of roughly 2.587 years, for which we estimate the amplitude to be of order one minute.

\ksec{Kepler-416 (KOI-1860)}
In the literature, we found only a nominal mass estimate of KOI-1860.01 (Kepler-416 b), provided by HL14.
The four transiting planets in this system form an 8:4:2:1 near-MMR chain.
We provide two solutions of similar fit quality, which are consistent in most parameters and differ mainly in the mass of the third planet and its orbital inclination with respect to the other planets. We select the adopted solution as the one that suggests a more co-planar structure and has a slightly better fit quality (by an amount equivalent to $\sim0.5\,\sigma$).

\ksec{Kepler-338 (KOI-1930)}
For this four-planet system, we found only nominal masses for planets KOI-1930.01 (Kepler-338 b) and KOI-1930.04 (Kepler-338 e) in HL14. Hence our masses for KOI-1930.02 (Kepler-338 c) and KOI-1930.03 (Kepler-338 d) are the first estimates we are aware of.
Our solution suggests masses of $5.9^{+2.1}_{-0.8}\,m_\oplus$ for KOI 1930.04, $5.4^{+0.3}_{-0.3}\,m_\oplus$ for KOI 1930.01, $5.5^{+0.3}_{-0.4}\,m_\oplus$ for KOI 1930.02, $6.1^{+0.4}_{-0.6}\,m_\oplus$ for KOI 1930.03, where the density of the innermost planet is significantly larger than the densities of the other planets in the system.

\ksec{Kepler-341 (KOI-1952)}
We found no literature mass estimates for this four-planet system.
We provide two solutions that agree on most of the fitted parameters, where the adopted one (chosen by its much better agreement with an N-body integration) suggests masses of $3.3^{+0.6}_{-0.6}\,m_\oplus$ for KOI 1952.03 (Kepler-341 b), $4.6^{+0.9}_{-0.9}\,m_\oplus$ for KOI 1952.01 (Kepler-341 c), $3.8^{+1}_{-1}\,m_\oplus$ for KOI 1952.02 (Kepler-341 d), $5.6^{+0.6}_{-0.6}\,m_\oplus$ for KOI 1952.04 (Kepler-341 e), and a co-planar structure with eccentricities of order a few percent.

\ksec{Kepler-85 (KOI-2038)}
In the literature, we found a nominal mass estimate for planet Kepler-85 c by HL14 and mass estimates for all planets by HL17.
We provide four different solutions, with the adopted one suggesting masses of $4.4^{+0.3}_{-0.4}\,m_\oplus$ for KOI 2038.01 (Kepler-85 b), $5.7^{+0.4}_{-0.5}\,m_\oplus$ for KOI 2038.02 (Kepler-85 c), $0.7^{+0.4}_{-0.3}\,m_\oplus$ for KOI 2038.03 (Kepler-85 d), $3.7^{+0.4}_{-0.5}\,m_\oplus$ for KOI 2038.04 (Kepler-85 e), consistent with the masses given by HL17. Our solution suggests a roughly co-planar, circular structure with no significant observed TbVs. In this case, due to the large number of different solutions, our confidence in the designation of a specific solution as adopted is not large; in order to constrain the system parameters, additional data would be required.
The planets in this system form a chain of three 3:2 MMRs. Kepler-85 c and Kepler-85 d have a super-period of roughly 130.8 days, and Kepler-85 d and Kepler-85 e have a super-period of approximately 136.4 days. This proximity of the super-periods gives rise to a non-pairwise TTV pattern, which we called Super-Mean-Motion-Resonance (SMMR) in paper I. Such an effect also occurs due to the proximity of other super-periods in this system; for example, planets b and d (innermost and third from inside) are affected by their proximity to the 2:1 MMR, with a super-period of roughly 114.2 days. This super-period is close to the 136.4 days super-period of planets d and e. This proximity of super-periods generates an SMMR of approximately 700 days timescale that affects the b-d-e triplet. Our best model estimates a total TTV of 7-8 minutes amplitude due to these SMMR effects.

\clearpage

\section{Summary and Future Prospects}\label{sec:Discussion}
This work completes our previous one (paper II), in which we analyzed two- and three-transiting planet Kepler systems light curves by fitting a model based on the analytic approach 
\texttt{AnalyticLC}, described in paper I. In the current work, we fit a model to light curves of Kepler four-, five- and six- transiting planet systems. The model takes into account the 3D interactions and therefore is sensitive to different types of transit variations in addition to typically modeled TTVs \citep[][ and others]{HaddenLithwick2014, HaddenLithwick2017, JontofHutter2021}. The usage of the entire light curve simultaneously yields, in many cases, tighter constraints on planetary mass, as shown in Figure~\ref{fig:MassErrorsHistogram1}, which allowed a very high yield of determined masses: of the 101 planets in systems with good solutions, significant mass was determined for 95 of them. This figure also shows that the contribution to new planetary masses in our two works together is most significant for small TTV amplitude planets. This is probably due to the difficulty of obtaining a good fit for systems with weak dynamical interactions that yield small TTVs. The global fit approach, which integrates the dynamical information from the entire light curve (TTVs, MMR-related TbVs, secular effects, SMMR etc.), succeeds in extracting additional information from the data.

In addition to planetary masses, these works investigated impact parameter variations. Because long-term TbVs result are attributed to orbital nodal precession, these signals are probes for mutual inclination within the system. We note here that in addition to these long-term variations, other TbV patterns arise from near-resonant terms and are automatically taken into account in our model; however, in the catalog we compiled here, we specifically address linear-trend TbVs. In paper II and this work, we compiled a list of TbV-displaying KOIs. We show the general spread of the TbV rates in Figure~\ref{fig:dbdtVsPmin1} and find that the typical rate (or, at least, the typical rate we can detect with a significance better than $2\,\sigma$) is of order 0.01$\,\rm yr^{-1}$. This limited sample of KOIs shows that rapid TbVs appear only in systems that host a planet orbiting the star with an orbital period of less than $\sim20$ days; this may be explained by the exchange of angular momentum in the process of planetary migration. This preliminary conclusion requires a larger sample to gain significance and a more detailed quantitative dynamical explanation.

We compute the number of orbits for a planet to cease transiting and find that this quantity typically decreases for increasing orbital periods; the trend is shown in Figure~\ref{fig:dbdtTimeScale1}. This dimensionless time scale is essential in planning long-term observation campaigns to search for Earth-like planets.

We foresee two types of future continuations of this work. One is extending our knowledge of the characteristics of individual planetary systems by using an analytic model based on \texttt{AnalyticLC}. This work and paper II concentrated on Kepler photometric data; however, large data sets of long-time-span RV observations are natural candidates for such an analytic approach, which is particularly efficient for long time spans. Another related direction would be to combine Kepler data with that from TESS \citep{RickerEtAl2010} and the upcoming PLATO  observations\citep{Rauer2014}, or RV data and astrometric data from GAIA \citep{GAIACollaborationEtAl2018}.
A second research avenue is a statistical analysis of the number of detected TbVs and their magnitudes to better understand the nature of planetary systems as a population. Previously, studies showed that an AMD-based model is more consistent with the number of detected Transit Duration Variations (TDVs) than a model based on a bi-modal distribution of planetary systems \citep{Millholland2021}. Using our significantly updated catalog of TbVs may allow further progress.

In paper I, we published the method and code implementation \texttt{AnalyticLC}, used in this study as the fitting model. We report here a few improvements to the code, mainly that it now enables fitting a joint model of photometry and RV more robustly than before. We also added the ability to incorporate the stellar quadruple moment in the dynamical calculations. The code is available at \textbf{\url{https://github.com/yair111/AnalyticLC}}.

\acknowledgments
This study was supported by the Helen Kimmel Center for Planetary Sciences and the Minerva Center for Life Under Extreme Planetary Conditions \#13599 at the Weizmann Institute of Science.

This paper includes data collected by the Kepler mission and obtained from the MAST data archive at the Space Telescope Science Institute (STScI). Funding for the Kepler mission is provided by the NASA Science Mission Directorate. STScI is operated by the Association of Universities for Research in Astronomy, Inc., under NASA contract NAS 5-26555. The Kepler data used in this work can be found in MAST: \dataset[10.17909/T9059R]{http://dx.doi.org/10.17909/T9059R}.

We thank the reviewers for their useful comments, which improved the quality of this work.

\clearpage



\appendix
\renewcommand\thesection{Appendix \Alph{section}}

\section{Tables}\label{appendix}

\renewcommand\thesubsection{\Alph{section}.\arabic{subsection}}

\subsection{Orbital and Physical Parameters}

\begin{table}
\caption{Physical and orbital elements of our solutions that pass all tests, including N-body matching, AIC improvement over a strictly-periodic model, and physically plausible planetary density. If multiple solutions were found, they are all presented; the adopted solution is shown in bold text. For a detailed description of the meaning of these parameters, see paper II. Only a part of the table is given here for illustration; the full table is given in the machine-readable file MRT1a.txt. The orbital elements written here estimate the so-called "free" values. We note, that in many cases inversion of the signs of either $I_x$ or $I_y$ or both can yield a solution with similar fit quality. There are two exceptional cases (also mentioned in the text): KOI-834 and KOI-707, for which inverting the signs of these angles would produce fits of quality worse by more than $2\,\sigma$. A common need is to use the instantaneous orbital elements, which are not fitted in our method. However, we provide a machine-readable table of instantaneous orbital elements which can be used to run an N-body integration for all the best-fitting solutions given in this work and our former work (paper II). These values are given in Table~\ref{tab:Instantaneous} and in a machine-readable form. \label{tab:PhysicalOrbital}}
\begin{turn}{90}
\begin{tabular}{c c c c c c c c c c c c } 
 \hline
id & Planet & KOI & Period [d] & $\mu~[10^{-6}]$  & $m~[m_{\oplus}]$  & $\rho_{\rm p}\,[{\rm g\,cm^{-3}}]$ & $R_{\rm p}~[R_{\oplus}]$  & $\Delta e_x$   & $\Delta e_y$   & $I_x~[^{\circ}]$   & $I_y~[^{\circ}]$ \\
[0.5ex] 
\hline\hline
\bf 1 & \bf Kepler-79 b & \bf 152.03 & \bf 13.4846 & \bf $\mathbf{17.77^{+1.3}_{-1.54}}$ & \bf $\mathbf{7.66^{+0.62}_{-0.73}}$ & \bf $\mathbf{1.08^{+0.12}_{-0.13}}$ & \bf $\mathbf{3.381^{+0.09}_{-0.09}}$ & \bf $\mathbf{0.0088^{+0.0024}_{-0.002}}$ & \bf $\mathbf{0.0161^{+0.0029}_{-0.0025}}$ & \bf $\mathbf{0}$ & \bf $\mathbf{0.869^{+0.087}_{-0.106}}$  \\ 
\bf  & \bf Kepler-79 c & \bf 152.02 & \bf 27.4022 & \bf $\mathbf{19.51^{+1.08}_{-1.3}}$ & \bf $\mathbf{8.4^{+0.55}_{-0.65}}$ & \bf $\mathbf{0.881^{+0.089}_{-0.098}}$ & \bf $\mathbf{3.73^{+0.1}_{-0.1}}$ & \bf $\mathbf{-0.0256^{+0.0039}_{-0.0049}}$ & \bf $\mathbf{-0.0047^{+0.0043}_{-0.0039}}$ & \bf $\mathbf{7.22^{+2.28}_{-1.83}}$ & \bf $\mathbf{-0.402^{+0.048}_{-0.043}}$  \\ 
\bf  & \bf Kepler-79 d & \bf 152.01 & \bf 52.0909 & \bf $\mathbf{21.9^{+1.11}_{-1.09}}$ & \bf $\mathbf{9.43^{+0.57}_{-0.6}}$ & \bf $\mathbf{0.138^{+0.013}_{-0.013}}$ & \bf $\mathbf{7.2^{+0.19}_{-0.19}}$ & \bf $\mathbf{0.0592^{+0.0051}_{-0.0053}}$ & \bf $\mathbf{-0.0445^{+0.0071}_{-0.0072}}$ & \bf $\mathbf{-6.54^{+2.77}_{-2.58}}$ & \bf $\mathbf{0.015^{+0.03}_{-0.031}}$  \\ 
\bf  & \bf Kepler-79 e & \bf 152.04 & \bf 81.0645 & \bf $\mathbf{5.63^{+0.96}_{-0.73}}$ & \bf $\mathbf{2.43^{+0.42}_{-0.33}}$ & \bf $\mathbf{0.243^{+0.055}_{-0.044}}$ & \bf $\mathbf{3.79^{+0.15}_{-0.15}}$ & \bf $\mathbf{-0.0383^{+0.0068}_{-0.0052}}$ & \bf $\mathbf{0.0076^{+0.0031}_{-0.0034}}$ & \bf $\mathbf{-5.76^{+2.26}_{-2.32}}$ & \bf $\mathbf{-0.885^{+0.011}_{-0.011}}$  \\ 
[1ex]
\hline
1 & Kepler-11 b & 157.06 & 10.304 & $5.07^{+0.49}_{-0.55}$ & $1.8^{+0.19}_{-0.22}$ & $1.5^{+0.19}_{-0.21}$ & $1.876^{+0.039}_{-0.037}$ & $0.0072^{+0.0043}_{-0.0045}$ & $-0.0379^{+0.007}_{-0.0066}$ & $0$ & $0.479^{+0.081}_{-0.068}$  \\ 
 & Kepler-11 c & 157.01 & 13.0249 & $6.53^{+0.67}_{-0.75}$ & $2.32^{+0.26}_{-0.3}$ & $0.491^{+0.066}_{-0.075}$ & $2.957^{+0.062}_{-0.059}$ & $-0.0251^{+0.0031}_{-0.0036}$ & $0.0191^{+0.0026}_{-0.0022}$ & $-3.92^{+0.86}_{-0.84}$ & $-0.59^{+0.14}_{-0.14}$  \\ 
 & Kepler-11 d & 157.02 & 22.6871 & $19.34^{+0.82}_{-0.83}$ & $6.88^{+0.44}_{-0.51}$ & $1.039^{+0.095}_{-0.107}$ & $3.309^{+0.069}_{-0.066}$ & $0.0381^{+0.0075}_{-0.0068}$ & $0.0017^{+0.0057}_{-0.0059}$ & $-5.5^{+0.98}_{-1.01}$ & $-0.021^{+0.056}_{-0.054}$  \\ 
 & Kepler-11 e & 157.03 & 31.9955 & $27.65^{+0.98}_{-0.88}$ & $9.83^{+0.59}_{-0.67}$ & $0.646^{+0.058}_{-0.064}$ & $4.368^{+0.091}_{-0.087}$ & $-0.0121^{+0.001}_{-0.001}$ & $0.0161^{+0.0014}_{-0.0013}$ & $-2.84^{+0.88}_{-0.87}$ & $1.0868^{+0.0076}_{-0.0067}$  \\ 
 & Kepler-11 f & 157.04 & 46.6857 & $5.69^{+0.47}_{-0.47}$ & $2.02^{+0.19}_{-0.21}$ & $0.648^{+0.077}_{-0.082}$ & $2.575^{+0.054}_{-0.051}$ & $0.0073^{+0.0015}_{-0.0014}$ & $-0.0204^{+0.0016}_{-0.0017}$ & $4.84^{+1.75}_{-1.65}$ & $0.493^{+0.025}_{-0.024}$  \\ 
 & Kepler-11 g & 157.05 & 118.379 & $68.23^{+1.08}_{-1.16}$ & $24.25^{+1.24}_{-1.53}$ & $3.06^{+0.26}_{-0.3}$ & $3.512^{+0.076}_{-0.073}$ & $-0.0181^{+0.0037}_{-0.0035}$ & $-0.0399^{+0.0095}_{-0.0089}$ & $4.27^{+2.12}_{-2.45}$ & $-0.033^{+0.033}_{-0.034}$  \\ 
2 & Kepler-11 b & 157.06 & 10.304 & $5.37^{+0.39}_{-0.4}$ & $1.91^{+0.17}_{-0.18}$ & $1.64^{+0.19}_{-0.2}$ & $1.853^{+0.041}_{-0.039}$ & $0.026^{+0.0048}_{-0.005}$ & $-0.0209^{+0.0073}_{-0.0076}$ & $0$ & $0.446^{+0.096}_{-0.086}$  \\ 
 & Kepler-11 c & 157.01 & 13.0249 & $8.23^{+0.89}_{-0.85}$ & $2.93^{+0.35}_{-0.35}$ & $0.614^{+0.082}_{-0.085}$ & $2.967^{+0.061}_{-0.059}$ & $-0.0192^{+0.0023}_{-0.0024}$ & $0.0221^{+0.0021}_{-0.0019}$ & $-3.59^{+0.83}_{-0.78}$ & $-0.696^{+0.088}_{-0.075}$  \\ 
 & Kepler-11 d & 157.02 & 22.6871 & $19.79^{+0.73}_{-0.72}$ & $7.03^{+0.43}_{-0.5}$ & $1.063^{+0.095}_{-0.107}$ & $3.309^{+0.069}_{-0.066}$ & $0.0221^{+0.0051}_{-0.0052}$ & $-0.0105^{+0.0068}_{-0.0069}$ & $4.35^{+0.81}_{-0.79}$ & $0.167^{+0.042}_{-0.044}$  \\ 
 & Kepler-11 e & 157.03 & 31.9956 & $31.34^{+2.19}_{-2.2}$ & $11.14^{+0.95}_{-1.03}$ & $0.731^{+0.079}_{-0.086}$ & $4.371^{+0.091}_{-0.087}$ & $-0.014^{+0.0012}_{-0.0011}$ & $0.01085^{+0.00077}_{-0.00082}$ & $3.3^{+0.65}_{-0.64}$ & $1.1182^{+0.0048}_{-0.0048}$  \\ 
 & Kepler-11 f & 157.04 & 46.6857 & $7.11^{+0.43}_{-0.42}$ & $2.53^{+0.2}_{-0.21}$ & $0.813^{+0.081}_{-0.089}$ & $2.572^{+0.053}_{-0.051}$ & $0.0082^{+0.0019}_{-0.002}$ & $-0.019^{+0.0017}_{-0.0017}$ & $-4.71^{+1.3}_{-1.3}$ & $-0.537^{+0.018}_{-0.019}$  \\ 
 & Kepler-11 g & 157.05 & 118.379 & $68.53^{+0.79}_{-0.78}$ & $24.36^{+1.22}_{-1.51}$ & $3.12^{+0.26}_{-0.29}$ & $3.496^{+0.073}_{-0.069}$ & $-0.0172^{+0.0055}_{-0.0052}$ & $0.0015^{+0.0077}_{-0.0089}$ & $5.52^{+1.6}_{-1.42}$ & $-0.077^{+0.018}_{-0.017}$  \\ 
[1ex]
\hline
\end{tabular}
\end{turn}
\end{table}

\clearpage

\subsection{Light-Curve Transit and MMR-Proximity Parameters}

\begin{table}
\caption{Light curve transit and MMR-proximity parameters of all analyzed planets. A part of the table is shown here for visualization; the full table is given in a machine-readable file MRT2a.txt. The parameters of the planets whose system has a valid dynamical solution appear first, as in Table \ref{tab:PhysicalOrbital}, with the adopted solutions marked using the ``Adopted" flag in the file and shown in bold text in the lines printed below. The parameters of the systems with no valid solutions appear afterward. We note that in many cases the impact parameter sign can be inverted, and in most systems, this would produce a fit with similar fit quality. The exceptional systems are KOI-834 and KOI-707 (see text, \S~\ref{subsec:OutOfPlaneForces}). We published a similar table for 2- and 3-planet systems in paper II, and we add it here for completeness after a slight correction to the values of $T,\tau,b$ in the file MRT2.txt. \label{tab:Observable}}
\begin{turn}{90}
\begin{tabular}{c c c c c c c c c c c c } 
 \hline
id & Planet & KOI & Period [d] & $\sigma_{\rm TTV}$ [min] & $J$ & $\Delta$ & S.P. [day]& $T~[\rm hr]$ & $\tau~[\rm min]$ & $b$ & $db/dt~[\rm yr^{-1}]$  \\
[0.5ex] 
\hline\hline
\bf 1 & \bf Kepler-79 b & \bf 152.03 & \bf 13.4846 & \bf 9.55 & \bf  & \bf  & \bf  & \bf $\mathbf{5.154^{+0.022}_{-0.021}}$ & \bf $\mathbf{7.87^{+0.13}_{-0.14}}$ & \bf $\mathbf{0.285^{+0.027}_{-0.033}}$ & \bf $\mathbf{0.003^{+0.0012}_{-0.0011}}$  \\ 
\bf  & \bf Kepler-79 c & \bf 152.02 & \bf 27.4022 & \bf 13 & \bf 2 & \bf 0.016 & \bf 853& $\mathbf{6.82^{+0.019}_{-0.019}}$ & \bf $\mathbf{11.08^{+0.13}_{-0.13}}$ & \bf $\mathbf{-0.217^{+0.025}_{-0.022}}$ & \bf $\mathbf{-0.0127^{+0.0015}_{-0.0015}}$  \\ 
\bf  & \bf Kepler-79 d & \bf 152.01 & \bf 52.0909 & \bf 6.83 & \bf 2 & \bf -0.05 & \bf 526& $\mathbf{8.1494^{+0.0099}_{-0.01}}$ & \bf $\mathbf{24.34^{+0.053}_{-0.054}}$ & \bf $\mathbf{0.012^{+0.023}_{-0.024}}$ & \bf $\mathbf{0.0092^{+0.0015}_{-0.0016}}$  \\ 
\bf  & \bf Kepler-79 e & \bf 152.04 & \bf 81.0645 & \bf 27.6 & \bf 3 & \bf 0.037 & \bf 721& $\mathbf{2.64^{+0.17}_{-0.19}}$ & \bf $\mathbf{62.07^{+9.02}_{-6.24}}$ & \bf $\mathbf{-0.9632^{+0.0051}_{-0.0051}}$ & \bf $\mathbf{0.0011^{+0.002}_{-0.0019}}$  \\ 
[1ex]
\hline
1 & Kepler-11 b & 157.06 & 10.304 & 6.91 &  &  & & $4.073^{+0.017}_{-0.017}$ & $4.025^{+0.043}_{-0.035}$ & $0.157^{+0.027}_{-0.022}$ & $-0.0141^{+0.0023}_{-0.0023}$  \\ 
 & Kepler-11 c & 157.01 & 13.0249 & 6.69 & 5 & 0.011 & 232& $4.451^{+0.028}_{-0.037}$ & $7.14^{+0.18}_{-0.15}$ & $-0.232^{+0.055}_{-0.054}$ & $0.0087^{+0.0038}_{-0.0037}$  \\ 
 & Kepler-11 d & 157.02 & 22.6871 & 5.57 & 2 & -0.13 & 87.9& $5.296^{+0.014}_{-0.014}$ & $9.003^{+0.04}_{-0.036}$ & $-0.011^{+0.03}_{-0.029}$ & $0.0147^{+0.0024}_{-0.0023}$  \\ 
 & Kepler-11 e & 157.03 & 31.9955 & 5.36 & 3 & -0.06 & 178& $3.922^{+0.023}_{-0.023}$ & $20.73^{+0.19}_{-0.19}$ & $0.7582^{+0.0035}_{-0.0035}$ & $-0.0017^{+0.0015}_{-0.0015}$  \\ 
 & Kepler-11 f & 157.04 & 46.6857 & 23.6 & 3 & -0.027 & 571& $6.086^{+0.062}_{-0.068}$ & $9.96^{+0.13}_{-0.12}$ & $0.438^{+0.02}_{-0.02}$ & $-0.0323^{+0.008}_{-0.008}$  \\ 
 & Kepler-11 g & 157.05 & 118.379 & 0.44 & 2 & 0.27 & 221& $9.364^{+0.036}_{-0.052}$ & $16.95^{+0.17}_{-0.13}$ & $-0.055^{+0.055}_{-0.057}$ & $-0.0019^{+0.00095}_{-0.0008}$  \\ 
2 & Kepler-11 b & 157.06 & 10.304 & 9.11 &  &  & & $4.06^{+0.017}_{-0.019}$ & $3.947^{+0.037}_{-0.034}$ & $0.142^{+0.031}_{-0.027}$ & $-0.0093^{+0.0032}_{-0.0032}$  \\ 
 & Kepler-11 c & 157.01 & 13.0249 & 6.39 & 5 & 0.011 & 231& $4.363^{+0.028}_{-0.028}$ & $7.142^{+0.097}_{-0.095}$ & $-0.264^{+0.034}_{-0.029}$ & $0.0217^{+0.0037}_{-0.0038}$  \\ 
 & Kepler-11 d & 157.02 & 22.6871 & 6.83 & 2 & -0.13 & 87.9& $5.3^{+0.015}_{-0.015}$ & $9.076^{+0.043}_{-0.041}$ & $0.089^{+0.022}_{-0.024}$ & $-0.0109^{+0.0031}_{-0.0032}$  \\ 
 & Kepler-11 e & 157.03 & 31.9956 & 5.73 & 3 & -0.06 & 178& $3.9^{+0.024}_{-0.023}$ & $21.12^{+0.22}_{-0.22}$ & $0.7648^{+0.0038}_{-0.004}$ & $-0.0035^{+0.0017}_{-0.0017}$  \\ 
 & Kepler-11 f & 157.04 & 46.6857 & 25.5 & 3 & -0.027 & 571& $6.015^{+0.055}_{-0.059}$ & $10.17^{+0.13}_{-0.11}$ & $-0.468^{+0.015}_{-0.016}$ & $0.0413^{+0.007}_{-0.0068}$  \\ 
 & Kepler-11 g & 157.05 & 118.379 & 0.44 & 2 & 0.27 & 221& $9.364^{+0.034}_{-0.037}$ & $17.08^{+0.15}_{-0.14}$ & $-0.127^{+0.03}_{-0.029}$ & $-0.00159^{+0.00054}_{-0.00062}$  \\ 
[1ex]
\hline
\end{tabular}
\end{turn}
\end{table}

\clearpage

\subsection{Stellar Parameters}

\begin{center}
\begin{longtable}{c c c c c c c c } 
\caption{Stellar parameters of the systems for which valid dynamical solutions were found (systems that appear in table \ref{tab:PhysicalOrbital}). The full table is given in the machine-readable file MRT3a.txt.} \\
\label{tab:Stellar} \\
 \hline
KOI & Star & KIC & $m~[m_{\odot}]$ & $R~[R_{\odot}]$ & $m,R$ source & $u_1$ & $u_2$  \\
[0.5ex] 
\hline\hline
152 & Kepler-79 & 8394721 & $1.294^{+0.044}_{-0.052}$ & $1.325^{+0.035}_{-0.035}$ & Berger et al. 2020 & 0.3201 & 0.305  \\ 
[1ex]
\hline
157 & Kepler-11 & 6541920 & $1.068^{+0.052}_{-0.065}$ & $1.071^{+0.022}_{-0.021}$ & Berger et al. 2020 & 0.4035 & 0.2622  \\ 
[1ex]
\hline
191 &  & 5972334 & $0.965^{+0.04}_{-0.049}$ & $0.912^{+0.031}_{-0.027}$ & Berger et al. 2020 & 0.4526 & 0.2327  \\ 
[1ex]
\hline
232 & Kepler-122 & 4833421 & $1.083^{+0.071}_{-0.073}$ & $1.244^{+0.037}_{-0.035}$ & Berger et al. 2020 & 0.3447 & 0.2916  \\ 
[1ex]
\hline
\end{longtable}
\end{center}

\clearpage

\subsection{Planets with Significant Impact Parameter Variations}

\begin{center}
\begin{longtable}{c c c c c c c }
\caption{Planets with significant impact parameter variations. First to be included in the table are systems with dynamical solutions, i.e., that also appear in Table \ref{tab:PhysicalOrbital}, with a $db/dt$ significant to $>2\,\sigma$. We show the values of $db/dt$ for the adopted solution only. The next group to be included in the table are systems without a dynamical solution but with a $db/dt$ significant to $>2\,\sigma$ for all our runs. The sign of $b_0$ is always positive for the innermost planet in the system. A negative value corresponds to a solution in which the transit occurs on the opposite hemisphere of the stellar disk than the innermost planet. We note that in many cases the impact parameter sign can be inverted, and in most systems, this would produce a fit with similar fit quality. The exceptional systems are KOI-834 and KOI-707 (see text, \S~\ref{subsec:OutOfPlaneForces}). The table is given in the machine-readable file MRT4a.txt. We published a similar table for 2- and 3-planet systems in paper II, and we add it here for completeness after a slight correction to the values of $b_0$ in the file MRT4.txt.} \\
\label{tab:SignificantDbdt} \\
 \hline
Planet & KOI & Period [d] & $N_{\rm pl}$ & Position & $b_0$ & $db/dt~[\rm yr^{-1}]$  \\
[0.5ex] 
\hline\hline
\bf Kepler-79 d & \bf 152.01 & \bf 52.0909 & \bf 4 & \bf 3 & \bf $\mathbf{0.012^{+0.023}_{-0.024}}$ & \bf $\mathbf{0.0092^{+0.0015}_{-0.0016}}$  \\ 
[1ex]
\hline
\bf Kepler-79 c & \bf 152.02 & \bf 27.4022 & \bf 4 & \bf 2 & \bf $\mathbf{-0.217^{+0.025}_{-0.022}}$ & \bf $\mathbf{-0.0127^{+0.0015}_{-0.0015}}$  \\ 
[1ex]
\hline
\bf Kepler-79 b & \bf 152.03 & \bf 13.4846 & \bf 4 & \bf 1 & \bf $\mathbf{0.285^{+0.027}_{-0.033}}$ & \bf $\mathbf{0.003^{+0.0012}_{-0.0011}}$  \\ 
[1ex]
\hline
\bf Kepler-11 c & \bf 157.01 & \bf 13.0249 & \bf 6 & \bf 2 & \bf $\mathbf{-0.335^{+0.013}_{-0.013}}$ & \bf $\mathbf{0.0161^{+0.002}_{-0.002}}$  \\ 
[1ex]
\hline
\bf Kepler-11 d & \bf 157.02 & \bf 22.6872 & \bf 6 & \bf 3 & \bf $\mathbf{0.037^{+0.034}_{-0.038}}$ & \bf $\mathbf{-0.0078^{+0.0026}_{-0.0025}}$  \\ 
[1ex]
\hline
\bf Kepler-11 e & \bf 157.03 & \bf 31.9955 & \bf 6 & \bf 4 & \bf $\mathbf{-0.7641^{+0.0036}_{-0.0037}}$ & \bf $\mathbf{0.0038^{+0.0017}_{-0.0016}}$  \\ 
[1ex]
\hline
\bf Kepler-11 f & \bf 157.04 & \bf 46.6858 & \bf 6 & \bf 5 & \bf $\mathbf{0.482^{+0.015}_{-0.014}}$ & \bf $\mathbf{-0.0448^{+0.0063}_{-0.0063}}$  \\ 
[1ex]
\hline
\bf Kepler-11 g & \bf 157.05 & \bf 118.379 & \bf 6 & \bf 6 & \bf $\mathbf{0.053^{+0.032}_{-0.034}}$ & \bf $\mathbf{0.00378^{+0.0008}_{-0.0008}}$  \\ 
[1ex]
\hline
\bf  & \bf 191.03 & \bf 0.70862 & \bf 4 & \bf 1 & \bf $\mathbf{0.056^{+0.027}_{-0.023}}$ & \bf $\mathbf{-0.0044^{+0.0022}_{-0.0022}}$  \\ 
[1ex]
\hline
\bf Kepler-487 c & \bf 191.04 & \bf 38.6511 & \bf 4 & \bf 4 & \bf $\mathbf{0.003^{+0.064}_{-0.063}}$ & \bf $\mathbf{-0.0069^{+0.003}_{-0.0028}}$  \\ 
[1ex]
\hline
\bf Kepler-122 c & \bf 232.01 & \bf 12.466 & \bf 5 & \bf 2 & \bf $\mathbf{0.269^{+0.022}_{-0.021}}$ & \bf $\mathbf{-0.013^{+0.0024}_{-0.0027}}$  \\ 
[1ex]
\hline
\bf Kepler-122 d & \bf 232.03 & \bf 21.5875 & \bf 5 & \bf 3 & \bf $\mathbf{-0.668^{+0.012}_{-0.013}}$ & \bf $\mathbf{0.0235^{+0.0038}_{-0.0036}}$  \\ 
[1ex]
\hline
\bf Kepler-122 e & \bf 232.04 & \bf 37.9965 & \bf 5 & \bf 4 & \bf $\mathbf{-0.069^{+0.034}_{-0.039}}$ & \bf $\mathbf{-0.0386^{+0.0056}_{-0.0058}}$  \\ 
[1ex]
\hline
\bf Kepler-122 f & \bf 232.05 & \bf 56.259 & \bf 5 & \bf 5 & \bf $\mathbf{0.102^{+0.043}_{-0.044}}$ & \bf $\mathbf{0.0111^{+0.0023}_{-0.0019}}$  \\ 
[1ex]
\hline
\bf Kepler-49 b & \bf 248.01 & \bf 7.20387 & \bf 4 & \bf 2 & \bf $\mathbf{-0.06^{+0.028}_{-0.025}}$ & \bf $\mathbf{-0.047^{+0.0056}_{-0.0052}}$  \\ 
[1ex]
\hline
\bf Kepler-49 c & \bf 248.02 & \bf 10.9127 & \bf 4 & \bf 3 & \bf $\mathbf{0.698^{+0.01}_{-0.01}}$ & \bf $\mathbf{-0.0146^{+0.0051}_{-0.0053}}$  \\ 
[1ex]
\hline
\bf Kepler-49 d & \bf 248.03 & \bf 2.57657 & \bf 4 & \bf 1 & \bf $\mathbf{0.041^{+0.015}_{-0.016}}$ & \bf $\mathbf{0.0344^{+0.0028}_{-0.0032}}$  \\ 
[1ex]
\hline
\bf Kepler-49 e & \bf 248.04 & \bf 18.5961 & \bf 4 & \bf 4 & \bf $\mathbf{-0.796^{+0.016}_{-0.017}}$ & \bf $\mathbf{0.0144^{+0.0073}_{-0.0067}}$  \\ 
[1ex]
\hline
\bf Kepler-26 b & \bf 250.01 & \bf 12.283 & \bf 4 & \bf 2 & \bf $\mathbf{0.003^{+0.044}_{-0.047}}$ & \bf $\mathbf{-0.0135^{+0.004}_{-0.0039}}$  \\ 
[1ex]
\hline
\bf Kepler-26 c & \bf 250.02 & \bf 17.2512 & \bf 4 & \bf 3 & \bf $\mathbf{0.8054^{+0.0066}_{-0.0068}}$ & \bf $\mathbf{-0.0072^{+0.0027}_{-0.0026}}$  \\ 
[1ex]
\hline
\bf Kepler-26 d & \bf 250.03 & \bf 3.54392 & \bf 4 & \bf 1 & \bf $\mathbf{0.318^{+0.031}_{-0.03}}$ & \bf $\mathbf{0.00914^{+0.00068}_{-0.00069}}$  \\ 
[1ex]
\hline
\bf Kepler-26 e & \bf 250.04 & \bf 46.8276 & \bf 4 & \bf 4 & \bf $\mathbf{0.8238^{+0.0097}_{-0.0093}}$ & \bf $\mathbf{0.0162^{+0.0016}_{-0.0018}}$  \\ 
[1ex]
\hline
\bf Kepler-169 e & \bf 505.01 & \bf 13.7671 & \bf 5 & \bf 4 & \bf $\mathbf{-0.588^{+0.019}_{-0.019}}$ & \bf $\mathbf{0.0107^{+0.0049}_{-0.005}}$  \\ 
[1ex]
\hline
\bf Kepler-169 c & \bf 505.02 & \bf 6.19548 & \bf 5 & \bf 2 & \bf $\mathbf{-0.131^{+0.055}_{-0.056}}$ & \bf $\mathbf{-0.058^{+0.014}_{-0.015}}$  \\ 
[1ex]
\hline
\bf Kepler-169 b & \bf 505.03 & \bf 3.25059 & \bf 5 & \bf 1 & \bf $\mathbf{0.207^{+0.047}_{-0.048}}$ & \bf $\mathbf{0.004^{+0.002}_{-0.0018}}$  \\ 
[1ex]
\hline
\bf Kepler-169 d & \bf 505.04 & \bf 8.34818 & \bf 5 & \bf 3 & \bf $\mathbf{-0.419^{+0.051}_{-0.049}}$ & \bf $\mathbf{0.025^{+0.01}_{-0.011}}$  \\ 
[1ex]
\hline
\bf Kepler-176 c & \bf 520.01 & \bf 12.7594 & \bf 4 & \bf 2 & \bf $\mathbf{0.04^{+0.064}_{-0.064}}$ & \bf $\mathbf{-0.0113^{+0.0047}_{-0.0052}}$  \\ 
[1ex]
\hline
\bf Kepler-176 b & \bf 520.02 & \bf 5.43312 & \bf 4 & \bf 1 & \bf $\mathbf{0.337^{+0.029}_{-0.028}}$ & \bf $\mathbf{0.0055^{+0.0014}_{-0.0014}}$  \\ 
[1ex]
\hline
\bf Kepler-176 e & \bf 520.04 & \bf 51.1656 & \bf 4 & \bf 4 & \bf $\mathbf{-0.686^{+0.031}_{-0.029}}$ & \bf $\mathbf{0.0093^{+0.0017}_{-0.0016}}$  \\ 
[1ex]
\hline
\bf Kepler-186 d & \bf 571.02 & \bf 13.343 & \bf 5 & \bf 3 & \bf $\mathbf{-0.059^{+0.064}_{-0.058}}$ & \bf $\mathbf{-0.051^{+0.016}_{-0.012}}$  \\ 
[1ex]
\hline
\bf Kepler-33 d & \bf 707.01 & \bf 21.7757 & \bf 5 & \bf 3 & \bf $\mathbf{-0.138^{+0.063}_{-0.046}}$ & \bf $\mathbf{0.0248^{+0.0051}_{-0.007}}$  \\ 
[1ex]
\hline
\bf Kepler-33 f & \bf 707.02 & \bf 41.0284 & \bf 5 & \bf 5 & \bf $\mathbf{0.212^{+0.03}_{-0.035}}$ & \bf $\mathbf{-0.0142^{+0.0028}_{-0.0021}}$  \\ 
[1ex]
\hline
\bf Kepler-33 e & \bf 707.03 & \bf 31.7847 & \bf 5 & \bf 4 & \bf $\mathbf{-0.172^{+0.032}_{-0.031}}$ & \bf $\mathbf{-0.0152^{+0.0052}_{-0.004}}$  \\ 
[1ex]
\hline
\bf Kepler-33 b & \bf 707.05 & \bf 5.66816 & \bf 5 & \bf 1 & \bf $\mathbf{0.222^{+0.057}_{-0.047}}$ & \bf $\mathbf{-0.00227^{+0.00077}_{-0.00069}}$  \\ 
[1ex]
\hline
\bf Kepler-221 c & \bf 720.01 & \bf 5.69059 & \bf 4 & \bf 2 & \bf $\mathbf{-0.196^{+0.024}_{-0.023}}$ & \bf $\mathbf{-0.0092^{+0.0031}_{-0.0036}}$  \\ 
[1ex]
\hline
\bf Kepler-221 d & \bf 720.02 & \bf 10.0416 & \bf 4 & \bf 3 & \bf $\mathbf{-0.6268^{+0.0103}_{-0.0092}}$ & \bf $\mathbf{0.0178^{+0.0024}_{-0.0024}}$  \\ 
[1ex]
\hline
\bf Kepler-221 e & \bf 720.03 & \bf 18.3698 & \bf 4 & \bf 4 & \bf $\mathbf{0.8328^{+0.0081}_{-0.0094}}$ & \bf $\mathbf{-0.0102^{+0.0019}_{-0.002}}$  \\ 
[1ex]
\hline
\bf Kepler-221 b & \bf 720.04 & \bf 2.7959 & \bf 4 & \bf 1 & \bf $\mathbf{0.092^{+0.022}_{-0.024}}$ & \bf $\mathbf{-0.0046^{+0.0014}_{-0.0014}}$  \\ 
[1ex]
\hline
\bf Kepler-235 b & \bf 812.01 & \bf 3.34023 & \bf 4 & \bf 1 & \bf $\mathbf{0.161^{+0.066}_{-0.058}}$ & \bf $\mathbf{0.0182^{+0.0067}_{-0.0057}}$  \\ 
[1ex]
\hline
\bf Kepler-235 c & \bf 812.04 & \bf 7.82498 & \bf 4 & \bf 2 & \bf $\mathbf{0.25^{+0.15}_{-0.21}}$ & \bf $\mathbf{-0.0312^{+0.0082}_{-0.008}}$  \\ 
[1ex]
\hline
\bf Kepler-27 b & \bf 841.01 & \bf 15.3353 & \bf 5 & \bf 2 & \bf $\mathbf{-0.268^{+0.035}_{-0.027}}$ & \bf $\mathbf{0.0119^{+0.0034}_{-0.005}}$  \\ 
[1ex]
\hline
\bf Kepler-27 c & \bf 841.02 & \bf 31.3303 & \bf 5 & \bf 3 & \bf $\mathbf{-0.039^{+0.044}_{-0.041}}$ & \bf $\mathbf{-0.011^{+0.0045}_{-0.0039}}$  \\ 
[1ex]
\hline
\bf  & \bf 841.05 & \bf 499.484 & \bf 5 & \bf 5 & \bf $\mathbf{-0.5^{+0.058}_{-0.049}}$ & \bf $\mathbf{-0.003^{+0.0013}_{-0.0023}}$  \\ 
[1ex]
\hline
\bf Kepler-31 b & \bf 935.01 & \bf 20.8603 & \bf 4 & \bf 2 & \bf $\mathbf{-0.426^{+0.017}_{-0.019}}$ & \bf $\mathbf{-0.0088^{+0.002}_{-0.002}}$  \\ 
[1ex]
\hline
\bf Kepler-31 c & \bf 935.02 & \bf 42.6341 & \bf 4 & \bf 3 & \bf $\mathbf{0.509^{+0.014}_{-0.013}}$ & \bf $\mathbf{0.0121^{+0.0017}_{-0.0016}}$  \\ 
[1ex]
\hline
\bf Kepler-31 d & \bf 935.03 & \bf 87.6471 & \bf 4 & \bf 4 & \bf $\mathbf{0.019^{+0.088}_{-0.085}}$ & \bf $\mathbf{-0.0116^{+0.0018}_{-0.0021}}$  \\ 
[1ex]
\hline
\bf  & \bf 935.04 & \bf 9.61734 & \bf 4 & \bf 1 & \bf $\mathbf{0.382^{+0.066}_{-0.059}}$ & \bf $\mathbf{0.0043^{+0.0015}_{-0.0015}}$  \\ 
[1ex]
\hline
\bf Kepler-84 b & \bf 1589.01 & \bf 8.72588 & \bf 5 & \bf 2 & \bf $\mathbf{0.123^{+0.04}_{-0.041}}$ & \bf $\mathbf{0.056^{+0.013}_{-0.015}}$  \\ 
[1ex]
\hline
\bf Kepler-84 c & \bf 1589.02 & \bf 12.883 & \bf 5 & \bf 3 & \bf $\mathbf{0.345^{+0.056}_{-0.07}}$ & \bf $\mathbf{-0.0375^{+0.0092}_{-0.0082}}$  \\ 
[1ex]
\hline
\bf Kepler-338 b & \bf 1930.01 & \bf 13.7271 & \bf 4 & \bf 2 & \bf $\mathbf{-0.277^{+0.018}_{-0.018}}$ & \bf $\mathbf{0.0079^{+0.0021}_{-0.0038}}$  \\ 
[1ex]
\hline
 \\ 
[1ex]
\hline
Kepler-89 b & 94.04 & 3.74318 & 4 & 1 & $0.151^{+0.042}_{-0.041}$ & $0.00199^{+0.00017}_{-0.00017}$  \\ 
[1ex]
\hline
Kepler-106 d & 116.04 & 23.98 & 4 & 3 & $-0.388^{+0.027}_{-0.022}$ & $0.0246^{+0.007}_{-0.007}$  \\ 
[1ex]
\hline
Kepler-107 e & 117.01 & 14.7492 & 4 & 4 & $0.2712^{+0.0094}_{-0.0108}$ & $-0.01194^{+0.00094}_{-0.00091}$  \\ 
[1ex]
\hline
Kepler-107 c & 117.02 & 4.90144 & 4 & 2 & $-0.26^{+0.021}_{-0.017}$ & $0.0216^{+0.0022}_{-0.0022}$  \\ 
[1ex]
\hline
Kepler-107 b & 117.03 & 3.18004 & 4 & 1 & $0.237^{+0.025}_{-0.033}$ & $-0.0123^{+0.002}_{-0.0024}$  \\ 
[1ex]
\hline
 & 750.03 & 14.5171 & 4 & 2 & $0.764^{+0.054}_{-0.066}$ & $-0.0177^{+0.006}_{-0.007}$  \\ 
[1ex]
\hline
Kepler-279 b & 1236.02 & 12.3098 & 4 & 1 & $0.152^{+0.027}_{-0.027}$ & $-0.0035^{+0.00026}_{-0.00025}$  \\ 
[1ex]
\hline
Kepler-279 d & 1236.03 & 54.4207 & 4 & 3 & $0.7483^{+0.0051}_{-0.005}$ & $-0.0236^{+0.0026}_{-0.0024}$  \\ 
[1ex]
\hline
 & 1236.04 & 98.3438 & 4 & 4 & $0.518^{+0.013}_{-0.013}$ & $0.00996^{+0.00093}_{-0.00085}$  \\ 
[1ex]
\hline
Kepler-304 b & 1557.01 & 3.2957 & 4 & 2 & $-0.263^{+0.028}_{-0.028}$ & $0.0358^{+0.0075}_{-0.0073}$  \\ 
[1ex]
\hline
Kepler-304 d & 1557.02 & 9.65348 & 4 & 4 & $-0.222^{+0.036}_{-0.032}$ & $0.035^{+0.0043}_{-0.0042}$  \\ 
[1ex]
\hline
Kepler-304 c & 1557.03 & 5.31597 & 4 & 3 & $-0.8811^{+0.0062}_{-0.0059}$ & $-0.0115^{+0.0021}_{-0.0021}$  \\ 
[1ex]
\hline
Kepler-305 d & 1563.04 & 16.7387 & 4 & 4 & $-0.869^{+0.012}_{-0.014}$ & $0.0179^{+0.0028}_{-0.0026}$  \\ 
[1ex]
\hline
Kepler-342 d & 1955.02 & 39.4575 & 4 & 4 & $0.119^{+0.057}_{-0.075}$ & $-0.0095^{+0.0024}_{-0.0024}$  \\ 
[1ex]
\hline
Kepler-342 e & 1955.03 & 1.64421 & 4 & 1 & $0.528^{+0.041}_{-0.045}$ & $-0.000134^{+6.3e-05}_{-6.2e-05}$  \\ 
[1ex]
\hline
Kepler-342 c & 1955.04 & 26.2351 & 4 & 3 & $-0.939^{+0.0049}_{-0.0047}$ & $0.007^{+0.0017}_{-0.002}$  \\ 
[1ex]
\hline
Kepler-402 d & 2722.04 & 8.92107 & 5 & 3 & $-0.229^{+0.051}_{-0.041}$ & $-0.0268^{+0.0039}_{-0.006}$  \\ 
[1ex]
\hline
Kepler-444 b & 3158.01 & 3.60011 & 5 & 1 & $0.0048^{+0.0055}_{-0.0033}$ & $0.0889^{+0.0017}_{-0.0019}$  \\ 
[1ex]
\hline
Kepler-444 c & 3158.02 & 4.54587 & 5 & 2 & $-0.5862^{+0.0038}_{-0.0043}$ & $-0.04202^{+0.00106}_{-0.00097}$  \\ 
[1ex]
\hline
Kepler-444 d & 3158.03 & 6.18945 & 5 & 3 & $-0.5063^{+0.0085}_{-0.009}$ & $-0.0119^{+0.0019}_{-0.0019}$  \\ 
[1ex]
\hline
Kepler-444 e & 3158.04 & 7.74343 & 5 & 4 & $-0.035^{+0.022}_{-0.023}$ & $-0.00844^{+0.00076}_{-0.00073}$  \\ 
[1ex]
\hline
Kepler-444 f & 3158.05 & 9.74048 & 5 & 5 & $-0.8364^{+0.003}_{-0.0028}$ & $-0.00463^{+0.00076}_{-0.00083}$  \\ 
[1ex]
\hline
\end{longtable}
\end{center}

\clearpage

\subsection{Systems with no Dynamical Solution}

\begin{center}
\begin{longtable}{c c c c } 
\caption{Systems that did not pass our tests and hence are not associated with a dynamical solution. For each system we provide the rejection reason. ``AIC" means that the strictly periodic model attains a better AIC \citep{Akaike1974} than the dynamical model (see \S\ref{sec:AlternativeModel}). ``N-body mismatch" means that the best fitting solution does not match an N-body integration to the level of $\sigma_{\rm Nbody}<1.5$ (see \S\ref{sec:Nbody}). ``High density" means that the median density is larger than $12\,{\rm g}\,{\rm cm}^{-3}$ by more than two error bars (see \S\ref{sec:Consistencty}).} \\
\label{tab:Reason} \\
 \hline
Planet & KOI & $N_{\rm pl}$ & Reason  \\
[0.5ex] 
\hline\hline
Kepler-89 & 94 & 4 & N-body mismatch  \\ 
[1ex]
\hline
Kepler-106 & 116 & 4 & N-body mismatch  \\ 
[1ex]
\hline
Kepler-107 & 117 & 4 & N-body mismatch  \\ 
[1ex]
\hline
Kepler-37 & 245 & 4 & AIC, N-body mismatch  \\ 
[1ex]
\hline
Kepler-152 & 416 & 4 & High density of innermost planet (KOI-416.03)  \\ 
[1ex]
\hline
Kepler-164 & 474 & 4 & AIC, N-body mismatch  \\ 
[1ex]
\hline
Kepler-167 & 490 & 4 & N-body mismatch, high density of third planet (KOI-490.04)  \\ 
[1ex]
\hline
Kepler-172 & 510 & 4 & N-body mismatch  \\ 
[1ex]
\hline
Kepler-197 & 623 & 4 & AIC  \\ 
[1ex]
\hline
Kepler-208 & 671 & 4 & AIC  \\ 
[1ex]
\hline
Kepler-215 & 700 & 4 & AIC, N-body mismatch  \\ 
[1ex]
\hline
Kepler-62 & 701 & 5 & AIC, high densities of planets 1 (KOI-701.02) and 4 (KOI-701.03)  \\ 
[1ex]
\hline
Kepler-220 & 719 & 4 & AIC, N-body mismatch, High density of the third planet (KOI-719.02)  \\ 
[1ex]
\hline
Kepler-224 & 733 & 4 & AIC  \\ 
[1ex]
\hline
Kepler-662 & 750 & 4 & AIC  \\ 
[1ex]
\hline
Kepler-251 & 907 & 4 & AIC, N-body mismatch  \\ 
[1ex]
\hline
Kepler-256 & 939 & 4 & AIC  \\ 
[1ex]
\hline
Kepler-265 & 1052 & 4 & AIC  \\ 
[1ex]
\hline
Kepler-758 & 1060 & 4 & AIC  \\ 
[1ex]
\hline
Kepler-763 & 1082 & 4 & AIC  \\ 
[1ex]
\hline
Kepler-271 & 1151 & 5 & AIC  \\ 
[1ex]
\hline
Kepler-279 & 1236 & 4 & N-body mismatch  \\ 
[1ex]
\hline
Kepler-286 & 1306 & 4 & AIC, N-body mismatch  \\ 
[1ex]
\hline
 & 1358 & 4 & AIC  \\ 
[1ex]
\hline
Kepler-292 & 1364 & 5 & AIC  \\ 
[1ex]
\hline
Kepler-296 & 1422 & 5 & AIC  \\ 
[1ex]
\hline
Kepler-299 & 1432 & 4 & AIC  \\ 
[1ex]
\hline
Kepler-304 & 1557 & 4 & N-body mismatch  \\ 
[1ex]
\hline
Kepler-305 & 1563 & 4 & N-body mismatch  \\ 
[1ex]
\hline
Kepler-306 & 1567 & 4 & AIC  \\ 
[1ex]
\hline
Kepler-87 & 1574 & 4 & N-body mismatch  \\ 
[1ex]
\hline
Kepler-324 & 1831 & 4 & N-body mismatch, high density of the innermost planet (KOI-1831.02).  \\ 
[1ex]
\hline
Kepler-342 & 1955 & 4 & N-body mismatch, high density of the innermost planet (KOI-1955.03).  \\ 
[1ex]
\hline
Kepler-352 & 2029 & 4 & AIC  \\ 
[1ex]
\hline
Kepler-1073 & 2055 & 4 & AIC  \\ 
[1ex]
\hline
Kepler-1130 & 2169 & 4 & AIC  \\ 
[1ex]
\hline
Kepler-374 & 2220 & 5 & AIC  \\ 
[1ex]
\hline
Kepler-402 & 2722 & 5 & AIC  \\ 
[1ex]
\hline
Kepler-403 & 2732 & 4 & AIC  \\ 
[1ex]
\hline
Kepler-1693 & 2871 & 4 & AIC  \\ 
[1ex]
\hline
Kepler-1388 & 2926 & 5 & AIC  \\ 
[1ex]
\hline
Kepler-444 & 3158 & 5 & N-body mismatch  \\ 
[1ex]
\hline
Kepler-1497 & 3429 & 4 & AIC  \\ 
[1ex]
\hline
Kepler-1581 & 4288 & 4 & AIC  \\ 
[1ex]
\hline
\end{longtable}
\end{center}
\clearpage

\subsection{Instantaneous Orbital Elements}

\begin{table}
\caption{Instantaneous coordinates, velocities, and orbital elements of the best-fitting parameters for the adopted solutions in this work and in paper II. The numbers are printed in a short version to allow the table to fit the page; the precise values are given in the file \texttt{InstantaneousOrbitalElements.csv} and in a \texttt{Matlab} format in \texttt{InstantaneousOrbitalElements.mat}. All parameters are given in MKS, apart from the epoch for which they are specified, which is given in days, with respect to Barycentric Kepler Julian Date (BKJD). The value of the gravitational constant used in our integrator is $G=6.674279896547770\times 10^{-11}$. The axis system is such that $yz$ forms the plane of the sky, and $x$ points towards the observer (transit occurs when $x>0$). 
\label{tab:Instantaneous}} 
\begin{turn}{90}
\begin{tabular}{c c c c c c c c c c c c c c c c c c c} 
 \hline
$x$ & $y$ & $z$ & $v_x$ & $v_y$ & $v_z$ & $m_*$ & $m_{\rm p}$ & $t$ & $a$ & $e$ & $\varpi$ & $I$ & $\Omega$ & $M$ & Pos. & KOI & Kepler- & $\sigma_{\rm Nbody}$  \\
[0.5ex] 
\hline\hline
-8.7e+09 & 5.1e+09 & 4.8e+08 & -6.2e+04 & -1.1e+05 & 3.5e+03 & 2.3e+30 & 1.5e+25 & 1.2e+02 & 1.1e+10 & 0.09 & 2.9 & 0.055 & 1.6 & 6.1 & 1 & 41.02 & 100 b & 0.8  \\ 
[1ex]
\hline
\end{tabular}
\end{turn}
\end{table}

\clearpage

\clearpage
\bibliography{Mybib.bib}{}
\bibliographystyle{aasjournal}


\end{document}